\documentclass[10pt,final,journal]{IEEEtran}
\usepackage[switch]{lineno} 
\usepackage{tabularx} 
\usepackage{booktabs}
\usepackage{multirow}
\usepackage{makecell}

\usepackage{stmaryrd}
\usepackage{graphicx}
\usepackage{amsthm} 
\usepackage{amsmath} 
\usepackage{bm}
\usepackage{graphicx}
\usepackage{epstopdf}
\usepackage{color}
\usepackage{float}
\usepackage[colorlinks=false,linkcolor=black,urlcolor=black,bookmarksopen=true, hidelinks]{hyperref}
\usepackage{bookmark}
\usepackage[flushleft]{threeparttable}
\usepackage{stackengine}
\usepackage{scalerel}
\usepackage{array}
\usepackage{stfloats}
\usepackage{lipsum}

\usepackage{xcolor,soul}

\usepackage{cite}

\ifCLASSINFOpdf
\else
\fi
\usepackage{amsfonts,amssymb}

\usepackage{amsmath}
\usepackage{algorithm}
\usepackage{algpseudocode}
\ifCLASSOPTIONcompsoc
 \usepackage[caption=false,font=normalsize,labelfont=sf,textfont=sf]{subfig}
\else
 \usepackage[caption=false,font=footnotesize]{subfig}
\fi
\begin{document}
%

\title{Robust Vehicle Positioning based on Multi-Epoch and Multi-Antenna TOAs in Harsh Environments}

\author{Xinyuan~An,
        Sihao~Zhao, \textit{Member, IEEE},
        Xiaowei~Cui,
        Gang~Liu,
        Mingquan~Lu
 \thanks{This work was supported by the National Key R\&D Program of China under Grant No. 2021YFA0716603. \textit{(Corresponding author: Xiaowei Cui)}}
\thanks{X. An, X. Cui and G. Liu are with the Department of Electronic Engineering,
	Tsinghua University, Beijing 100084, China (e-mail: anxinyuan1983@163.com; cxw2005@tsinghua.edu.cn; liu\_gang@tsinghua.edu.cn).}
\thanks{S. Zhao is with NovAtel,  Autonomy \& Positioning division of
Hexagon, Calgary, T3K 2L5, Canada (e-mail: zsh01@tsinghua.org.cn).}
\thanks{M. Lu is with the Department of Electronic Engineering,
	Beijing National Research Center for Information Science and Technology, Tsinghua University, Beijing 100084, China (e-mail: lumq@tsinghua.edu.cn).}
}

\markboth{}%
{Shell \MakeLowercase{\textit{et al.}}: Bare Demo of IEEEtran.cls for IEEE Journals}
%




\maketitle

\begin{abstract}
For radio-based time-of-arrival (TOA) positioning systems applied in harsh environments, obstacles in the surroundings and on the vehicle itself will block the signals from the anchors, reduce the number of available TOA measurements and thus degrade the localization performance.
Conventional multi-antenna positioning technique requires a good initialization to avoid local minima, and suffers from location ambiguity due to insufficient number of TOA measurements and/or poor geometry of anchors at a single epoch.
In this paper, taking advantage of the multi-epoch and multi-antenna (MEMA) TOA measurements bridged by inter-epoch constraints to utilize more information and improve the geometry of visible anchors, we propose a new positioning method, namely MEMA-TOA method.
A new initialization method based on semidefinite programming (SDP), namely MEMA-SDP, is first designed to address the initialization problem of the MEMA-TOA method. Then, an iterative refinement step is developed to obtain the optimal positioning result based on the MEMA-SDP initialization.
 We derive the Cram\'er-Rao lower bound (CRLB) to analyze the accuracy of the new MEMA-TOA method theoretically, and show its superior positioning performance over the conventional single-epoch and multi-antenna (SEMA) localization method. Simulation results in harsh environments demonstrate that i) the new MEMA-SDP provides an initial estimation that is close to the real location, and empirically guarantees the global optimality of the final refined positioning solution, and ii) compared with the conventional SEMA method, the new MEMA-TOA method has higher positioning accuracy without location ambiguity, consistent with the theoretical analysis.
\end{abstract}


%
\IEEEpeerreviewmaketitle

\section{Introduction}\label{Introduction}
%
%
%
%
\IEEEPARstart{R}{E}AL-time, continuous and accurate positions of the vehicles are pervasively needed in the intelligent transportation systems (ITS). In addition to the traditional guidance methods based on the prearranged markers on fixed routes, \cite{Le-Anh2006AGV,Chen2018AGV, mohamed2019survey}, more advanced positioning techniques, such as inertial \cite{Chatfield1997INS,Wang2013INS,LIwei2018RTKINS}, visual\cite{Taketomi2017SLM}, laser\cite{cadena2016past, Durrant2006SLAM,Hata2016Lidar} and radio based system are applied to ITS to realize vehicle guidance with easy route customization and low maintenance cost.

Due to high positioning accuracy, simple infrastructure, flexible deployment and independence of the navigation routes, radio-based positioning systems have been widely studied and applied \cite{Zamora2021AGV}. Measurements at the receiving antennas installed on the vehicle with respect to multiple anchors are used to estimate the location of the vehicle. These measurements include received signal strength (RSS), direction-of-arrival (DOA), time-of-flight (TOF), and time-of-arrival (TOA)  \cite{Sahinoglu2008UWBtheoretical,Shen2010UWBfundamental,zhao2021optimal}. 
The TOA-based technology has been used in the global navigation satellite systems (GNSSs) \cite{zhao2014analysis,zhao2021opt,fan2021two}, and also in many positioning systems and applications based on ultra-wide band (UWB) signal\cite{zhang2006UWBTOA,Djaja2018synchronized,Shi2020BLAS,Matsumoto2011UWBTOA,Sidorenko2020UWBTOA,Tiemann2016ATLAS}.

The number and quality of TOA measurements of the line-of-sight (LOS) paths are decisive factors for accuracy and availability of a TOA-based  positioning system.
In open areas or environments with few obstacles, the receiver with a single antenna can obtain sufficient TOA measurements with a high quality and a uniform anchor geometry, which ensure a good positioning performance. 

However, in a number of practical ITS applications such as urban automated guided vehicles, unmanned cargo ports and intelligent warehouses, there will be obstacles such as hoisting facilities, buildings and containers that may seriously impact the signal transmission and reception\cite{Soatti2018ICP}. The vehicle body itself will also affect the signal reception. For example, the intelligent transport vehicles at a modern cargo port usually have no cab and adopts a flat structure to load a large volume of goods. Thereby, the receiving antenna cannot be installed at the highest point of the vehicle, nor placed on top of a high-rise pole, which may swing when the vehicle moves and degrade the positioning accuracy. As a result, the receiving antenna can only be placed on the flat body of the vehicle, leading to severe signal blockage by the goods and the vehicle body. In such cases, the traditional single antenna positioning technique cannot guarantee the availability and accuracy of the positioning due to limited quality and quantity of available TOA measurements \cite{Wymeersch2009CP}, \cite{Nguyen2015LSCP}.


Aiming to solve the problem of insufficient measurements, multi-user collaborative positioning has become popular in recent years \cite{ Huang2016CP}, \cite{Lobo2019CL}. This technology relies on the stable operation of the inter-user wireless communication link \cite{Wang2021CP}, which requires a complex communication protocol and is susceptible to harsh environments such as a cargo port. In addition to the collaborative positioning system, a multi-antenna positioning system is proposed in \cite{An2020DMA}. Taking the advantage of the spatial diversity between multiple antennas, the availability and accuracy of the positioning system are improved. However, the cost function of the maximum likelihood estimator (MLE) to this multi-antenna positioning problem is nonlinear and non-convex, similar to the single-antenna case. The iterative algorithm to solve this problem may not converge or may be trapped at a local minimum if it has an inaccurate initial guess \cite{Zhao2021SDP}. In addition, location ambiguities may appear and lead to unacceptable errors when there are insufficient measurements and/or a poor geometry of anchors \cite{kaplan2005GPS,Kannan2010Ambiguity, Patrik2012Multilateration}.

Some studies in the literature about rigid body localization also discussed similar positioning problem by using the measurements from multiple sensors (or antennas) at different positions on the target to be localized \cite{Chen2015RLB_DAC,Jiang2018RBL_SDP,Jiang2019RBL_SDP2}. 
However, in the harsh environment described above, these methods are not applicable due to insufficient number of available measurements at a single epoch. Furthermore, their techniques adopt the distance or TOF measurements without the clock bias. The measurements with clock bias were handled in \cite{Ke2020RBL_clock}, in which the position and attitude were simultaneously estimated by relaxing the original problem using semidefinite relaxation (SDR). However, it still suffers from the problem of lacking measurements in harsh environments. Moreover, the result is sub-optimal to the original problem, and cannot be directly used for high-precision positioning applications. 

In order to position in harsh environments with frequent blockages, some researchers developed positioning techniques based on multi-system integration, such as combining the radio positioning system with one or several sensors to improve the positioning performance \cite{Farrell2008Aided,Vikram2017GPSvision,ding2020UWB_Vision_AGV,Laftchiev2015IMU,Wang2018MAINS}. 
To fuse the data from different systems such as GNSS and inertial navigation system (INS), a group of filtering methods, such as the extended Kalman filter (EKF) among the others, are frequently-used in these studies. Utilizing the information of previous epochs, the average positioning accuracy is improved in these integrated systems. In order to mitigate the accuracy degradation caused by outliers in the measurements, additional methods or robust filters \cite{Fang2022robustEKF,Chughtai2020EKF} that can detect or eliminate the outliers are needed to ensure the accuracy and robustness of such systems. Another way to use fused data for localization is to solve a nonlinear optimization problem by utilizing multiple past measurements to estimate the state at the current epoch \cite{Brembeck2019MHE,Zhao2016MHE}. However, whether iterative method or filtering method is used, it may suffer from the problem of inaccurate initialization.

In this paper, we propose a new positioning method based on multi-epoch and multi-antenna (MEMA) TOA measurements, namely MEMA-TOA method, for vehicle positioning in harsh environments. Different from the conventional localization methods, which adopt only single-epoch TOA measurements, we utilize TOAs from multiple recent epochs by introducing inter-epoch constraints on the changes of position and attitude. In this way, we are able to remove the location ambiguity and improve the geometry of the observed anchors. In the new MEMA-TOA method, we develop an initialization method based on semidefinite programming (SDP), called MEMA-SDP. In MEMA-SDP, the original positioning problem is approximated by a convex problem, which has a global optimal solution. Moreover, we develop a refinement method to obtain the optimal solution iteratively from the initial position from the MEMA-SDP. To analyze the positioning accuracy of the new MEMA-TOA method, we derive the Cram\'er-Rao lower bound (CRLB) and show its superior accuracy over the conventional single-epoch method. We conduct numerical simulations to evaluate the performance of the proposed method in harsh environments. Simulation results demonstrate that the new MEMA-SDP provides the initial values that are close to the real locations, and ensures the accuracy of the final refined positioning solution. Compared with the conventional single-epoch method, the new MEMA-TOA method successfully removes the location ambiguity and improves the positioning accuracy and robustness, consistent with the theoretical analysis. 

The main contributions of this paper are summarized as follows:

1) TOA measurements from MEMA and inter-epoch constraints are introduced to solve the location ambiguity caused by insufficient measurements at a single epoch for vehicle positioning under harsh environments.

2) The proposed MEMA-TOA method estimates the high-precision position and attitude of the vehicle through SDR-based initialization and iterative refinement, without requiring a priori initial value.

The remainder of the article is organized as follows. Section \ref{problem} gives the scenario and formulation of the vehicle positioning problem. The proposed method is developed in Section \ref{method}. Section \ref{performance} elaborates the performance of the multi-epoch and multi-antenna positioning by deriving the corresponding CRLB. Section \ref{simulation} presents the simulation results. And finally, the last section concludes the paper.

Main notations are summarized in Table \ref{table_notation}.
\begin{table}[!t]
	\caption{Notation List}
	\label{table_notation}
	\centering
	\begin{tabular}{l p{5cm}}
		\toprule
		lowercase $x$&  scalar\\
		bold lowercase $\boldsymbol{x}$ & vector\\
		bold uppercase $\bm{X}$ & matrix\\		
		$\hat{x}$, $\hat{\boldsymbol{x}}$ & estimate of a variable\\
		$\tilde{x}$, $\tilde{\boldsymbol{x}}$ & noisy version of a variable\\
		$\bar{x}$, $\bar{\boldsymbol{x}}$  & approximation of a variable\\
		$\Vert \boldsymbol{x} \Vert$ & Euclidean norm of a vector\\	
		$\bm{X}^T$, $\bm{X}^{-1}$ & matrix transpose and inverse, respectively\\
		$\operatorname{diag}(\cdot)$ & diagonal matrix with the elements inside along the diagonal\\
		$\operatorname{blkdiag}(\cdot) $& block diagonal matrix with the matrices inside along the diagonal\\
		$ \mathrm{tr}(\cdot)$, $\mathrm{det}(\cdot)$, $\mathrm{rank}(\cdot)$& trace, determinant and rank of a matrix, respectively\\
		$\bm{X}_{m\times n}$ & matrix with  $m$ rows and $n$  columns\\
		$[\bm{X}]_{m,n}$ & element at the $m$-th row and  $n$-th column of a matrix\\ $[\boldsymbol{x}]_{m:n}$  & the  $m$-th to  $n$-th elements of a vector\\ 
		$[\boldsymbol{x}]_{m}$ &the $m$-th element of vector $ \boldsymbol{x} $\\
		$\mathrm{vec}(\bm{X})$  &the vectorization of a matrix\\
		$\otimes$ & Kronecker product\\     	
		$M$ & number of anchors\\
		$N$ & number of antennas\\
		$K$ & number of epochs\\
		$i$, $ j$, $k$ & indices of antennas, anchors and epochs, respectively\\
		$\bm{I}_M$ & $M\times M$ identity matrix\\
		$\bm{0}_{M\times N}$ & $M\times N$ zero matrix\\
		$\boldsymbol{1}_{M}$& $M$-element vector with all-one elements\\
		$\boldsymbol{\theta}$, $\boldsymbol{\Theta}$ &  parameter vector and the collective form\\	
		$\mathrm{b}_{(k)}$ & body frame at epoch $k$ \\
		$\mathrm{n}$ &  navigation frame \\	
		$\bm{R}_{(k)}$&	 rotation matrix of transformation from frame $\mathrm{b}_{(k)}$ to frame $\mathrm{n}$ \\
		$\boldsymbol{p}_{i(k)}$ & unknown position vector of antenna $i$\\
		$\boldsymbol{p}^{(j)}_{(k)}$ & known position vector of anchor $j$\\
		$\boldsymbol{p}_{\mathrm{c}(k)}$ &  unknown position vector of the vehicle\\		
		$\delta t_{(k)}$ & clock bias between antennas and anchor system\\
		$\rho_{i(k)}^{(j)}$ & TOA measurement for antenna $i$ from the $j$-th visible anchor at epoch $k$ \\
		$r_{i(k)}^{(j)}$ & distance from antenna $i$ to its $j$-th visible anchor at epoch $k$ \\
		$h$ & height of the vehicle \\
		$ \phi$, $\gamma$, $\psi$ &  pitch, roll and yaw angle of the body, respectively \\ 
			$\mathrm{s}_{\phi}$, $\mathrm{s}_{\gamma}$, $\mathrm{s}_{\psi}$, $\mathrm{c}_{\phi}$, $ \mathrm{c}_{\gamma}$, $\mathrm{c}_{\psi}$ & $ \operatorname{sin}\phi$, $\operatorname{sin}\gamma$, $\operatorname{sin}\psi$, $ \operatorname{cos}\phi$, $\operatorname{cos}\gamma$ and $\operatorname{cos}\psi$, respectively \\ 
		$\mathrm{card}(\mathcal{B})$ & cardinal number of elements in set  $\mathcal{B}$\\
		$\mathtt{F}$ &  Fisher information matrix (FIM)\\
		$\mathtt{J}$ &   cost function of the optimization problem\\
		\bottomrule
	\end{tabular}
\end{table}

\section{Problem Formulation} \label{problem}
\subsection{System Settings}
\begin{figure}
	\centering
	\includegraphics[width=0.99\linewidth]{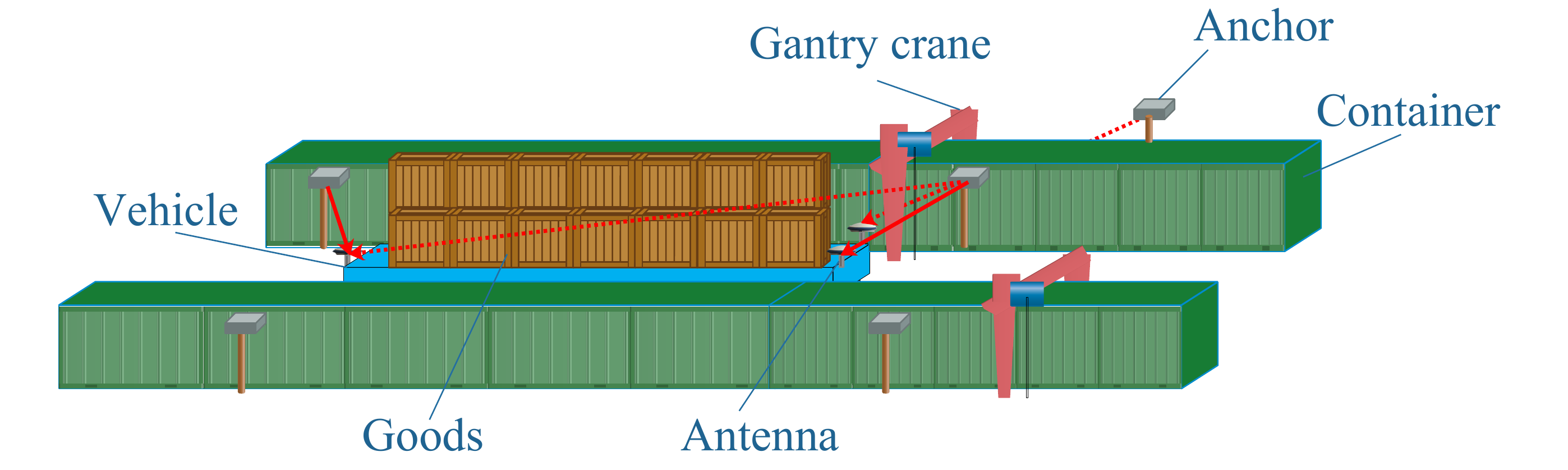}
	\vspace{-0.4cm}
	\caption{A typical vehicle localization scenario in an unmanned cargo port. The blue, flat and long cuboid represents the intelligent transport vehicle to be located, and the wooden cuboid represents the loaded goods.
	The gray cuboids are the anchors that transmit positioning signals. The receiving antennas are mounted on the edges of the vehicle. The containers, gantry cranes and goods may block the propagation of signals. The red solid line indicates the signal that can be received, while the red dotted line indicates the blocked signal.}
	\label{fig:scene_example}
	\vspace{-0.3cm}
\end{figure}

Fig. \ref{fig:scene_example} illustrates a typical vehicle localization scenario in an unmanned cargo port. The vehicles transporting goods are navigated with the help of the radio positioning system. There are hoisting facilities and containers in the port as well as goods loaded on the vehicle, which may block the propagation of the positioning signals. Anchors with fix positions are synchronized to a common clock source in various ways, such as wired connections between anchor nodes and wireless synchronization schemes \cite{Kegen2009wired,Djaja2018synchronized,Shi2020BLAS,Zhao2021TOA}. Multiple receiving antennas with known local positions relative to the origin of the body frame are mounted on the vehicle. The antennas have the same timing source, hence they have the same clock bias with respect to the anchors. In addition, an auxiliary sensor is installed on the body to provide the changes of the vehicle position and attitude between successive epochs. 

There are $M$ anchors and $N$  antennas in this positioning system. We take the output time of each set of TOA measurements at the antennas as an epoch. Let $\mathcal{N}$ be the set of antennas with $\mathrm{card(\mathcal{N})}=N$, $\mathcal{M}$ be the set of anchors with $\mathrm{card(\mathcal{M})}=M$, and $\mathcal{M}_{i(k)}$ be the set of visible anchors of antenna $i$ at epoch $k$ with $\mathrm{card}\left(\mathcal{M}_{i(k)}\right)=M_{i(k)}$ ,   $\mathcal{M}_{i(k)}\subseteq\mathcal{M}$. 

Without loss of generality, we assume that the vehicle runs on a flat area without changing the height, roll angle and pitch angle. Therefore, the height $ h $, roll angle $ \gamma $ and pitch angle $ \phi  $ are regarded as known constants.

\subsection{Reference Frame}\label{reference frame}
The known positions of the anchors and the unknown position of the vehicle are all expressed in the navigation frame, denoted by frame $\mathrm{n}$. 
We define the body frame with the center of the vehicle platform as the origin, denoted by frame $\mathrm{b}$, which changes with the movement of the vehicle and is expressed as $\mathrm{b}_{(k)}$ at epoch $k$. 




We denote the unknown position of the reference point on the vehicle by $\bm{p}_{\mathrm{c}(k)}$, the unknown position of antenna $i$ by $\bm{p}_{i(k)}$, and the known position of anchor $j$ by $\bm{p}_{(k)}^{(j)}$, respectively in frame $\mathrm{n}$. The subscript ``$(k)$'' represents the epoch index $k$. For simplicity, we do not specify frame $\mathrm{n}$ in the notations. Since the known local positions of the antennas in frame $\mathrm{b}_{(k)}$ do not change when the vehicle moves, we denote it by $\bm{l}_i$ without specifying frame $\mathrm{b}$ and epoch $k$. 


The relation between the antenna position $\bm{p}_{i(k)}$ in frame $\mathrm{n}$, the position of the reference point of the vehicle $\bm{p}_{\mathrm{c}{(k)}}$ in frame $\mathrm{n}$, and the antenna position $\bm{l}_{i}$ in frame $\mathrm{b}$ is \cite{Farrell2008Aided}
\begin{equation}\label{eq:pi}
	\bm{p}_{i(k)}=\bm{p}_{\mathrm{c}{(k)}}+\bm{R}_{(k)}\bm{l}_{i} \text{, } i \in\mathcal{N}
	\text{,}
\end{equation}
where $\bm{R}_{(k)}$ is the rotation matrix from frame $\mathrm{b}_{(k)}$ to frame $\mathrm{n}$. See Appendix \ref{Appendix_VEC} for the detailed definition of the rotation matrix. 

In this two-dimensional (2D) case,  $\bm{R}_{(k)}$  is a function of the yaw angle, denoted by $\psi_{(k)}$, and $ \bm{p}_{\mathrm{c}{(k)}}= \left[ x_{(k)},\ y_{(k)},\ h\right]^T $. The parameter to be estimated, denoted by $\bm{\theta}_{(k)}$, is 
\begin{align}
    \bm{\theta}_{(k)}= \left[
	x_{(k)},\ y_{(k)},\ \psi_{(k)} \right]^T.\nonumber
\end{align}

\subsection{TOA Measurement}\label{lasproblem}
In this paper, we consider only the TOA measurements from LOS paths and ignore the non-LOS measurements, which can be identified and eliminated \cite{groves2010novel, Xu2020NLOS, Yu2019NLOS,Li2005NLOS}. Following the model of the singe-epoch and single-antenna TOA measurement in \cite{kaplan2005GPS}, we extend it to the multi-epoch and multi-antenna case as
\begin{equation}\label{eq:rhoij}
\rho_{i(k)}^{(j)}=\Vert\bm{p}_{(k)}^{(j)}-\bm{p}_{i(k)}\Vert+\delta t_{(k)}+\varepsilon^{(j)}_{i(k)}
,\; i \in\mathcal{N},\ j\in\mathcal{M}_{i(k)},
\end{equation}
where $ \rho_{i(k)}^{(j)} $ is the TOA measurement for antenna $i$ from its $j$-th visible anchor at epoch $k$, $\delta t_{(k)}$ is the common clock bias between all the synchronous antennas and all the synchronous  anchors at epoch $k$, and $\varepsilon^{(j)}_{i(k)}$ is the measurement noise,
which is independent and identically distributed Gaussian white noise, i.e., $\varepsilon^{(j)}_{i(k)}\sim\mathrm{N}(0,\sigma^2)$.

\subsection{Inter-epoch Position and Attitude Change Constraint}
The changes of the vehicle position and attitude relative to the previous epochs can be measured by sensors such as the inertial measurement unit (IMU) and odometer in practice
\cite{mohamed2019survey}. We employ this information as a constraint for the vehicle position.

We model the inter-epoch position and attitude change constraint as
\begin{equation}\label{eq:odo1}	
	\Delta\tilde{\bm{\theta}}_{(k,k-1)}^{\mathrm{b}_{(k-1)}}=	\Delta \bm{\theta}_{(k,k-1)}^{\mathrm{b}_{(k-1)}}+\bm{\varepsilon}_{\mathrm{IP}(k)} \text{, }
\end{equation}
where $\Delta\bm{\theta}_{(k,k-1)}^{\mathrm{b}_{(k-1)}}=\left[ 
	\Delta x_{(k,k-1)}^{\mathrm{b}_{(k-1)}},\		\Delta y_{(k,k-1)}^{\mathrm{b}_{(k-1)}},\
		\Delta \psi_{(k,k-1)}^{\mathrm{b}_{(k-1)}}
\right]^T $ is the position and attitude change at epoch $ k $ relative to epoch $(k-1)$ in frame $\mathrm{b}_{(k-1)}$,  $\Delta\tilde{\bm{\theta}}_{(k,k-1)}^{\mathrm{b}_{(k-1)}}$ is the measurement of  $\Delta\bm{\theta}_{(k,k-1)}^{\mathrm{b}_{(k-1)}}$. A number of sensors such as odometer, magnetometer or encoder disk can be adopted to provide the measurements of inter-epoch position and attitude change constraint. We model the measurement errors as Gaussian noises since we do not specify any particular sensor and the characteristics of the sensor are not exactly known. $\bm{\varepsilon}_{\mathrm{IP}(k)}=\left[ 
	\varepsilon_{x_{(k)}},\ \varepsilon_{y_{(k)}},\ \varepsilon_{\psi_{(k)}}
\right]^T $ denotes the noise vector, in which the noises of the inter-epoch position change are modeled as independent and identically distributed Gaussian random variables as $\varepsilon_{x_{(k)}},\varepsilon_{y_{(k)}} \sim\mathrm{N}(0,\sigma_p^2)$, and the inter-epoch yaw angle change noise is modeled as a Gaussian random variable as $\varepsilon_{\psi_{(k)}} \sim\mathrm{N}(0,\sigma_\psi^2)$ \cite{Chen2020odo}.  

\subsection{Positioning Problem} \label{problem_description}
The positioning problem is to estimate $ \bm{\theta}_{(k)}$ at each epoch by using the multi-epoch and multi-antenna TOA and inter-epoch constraints. The difficulties of solving this problem lie in the initialization and the location ambiguity removal.

Firstly, the positioning problem is a nonlinear and non-convex optimization problem. Solving this problem based on the iterative MLE achieves the asymptotic optimality, on the condition that it has an accurate initial guess to start the iteration. Otherwise, the iteration may not converge or will be trapped at a local minimum \cite{Jiang2019RBL_SDP2,Zhao2021SDP}.

In addition, ambiguity of locations may appear when the number of TOA measurements is insufficient or the geometry of the observed anchors is inappropriate in harsh environments. These ambiguous solutions are also feasible solutions to the problem, but will lead to unacceptable errors if they are mistaken as the positioning results. 
For a regional positioning system, the difference between feasible solutions may be very small, sometimes only at meter level or even decimeter level. Therefore, it is difficult to identify the wrong solutions \cite{Kannan2010Ambiguity, Patrik2012Multilateration}.  
 
  \begin{figure}
  	\centering
  	\includegraphics[width=0.90\linewidth]{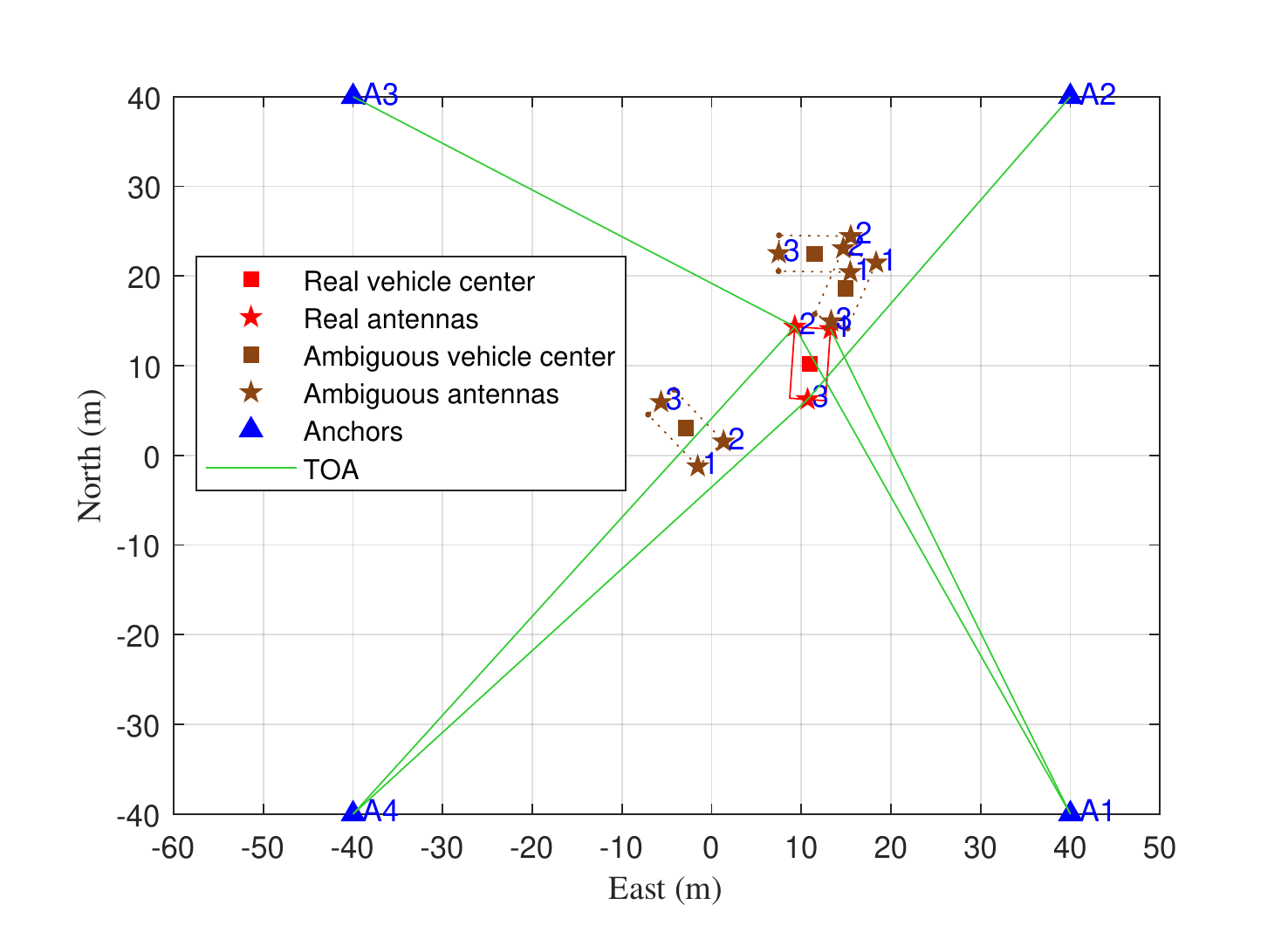}
  	\vspace{-0.2cm}
  	\caption{An example of location ambiguity in a multi-anchor and multi-antenna scenario. The green lines represent the TOA measurements between the antennas and their visible anchors. The red square and red stars represent the real locations of the vehicle center and its antennas, while the brown ones are the ambiguous solutions. These ambiguous location solutions are caused by the poor relative geometry between the antennas and their visible anchors.}
  	\label{fig:ambiguity}
  	\vspace{-0.3cm}
  \end{figure}

Fig. \ref{fig:ambiguity} gives an example of location ambiguity. Four ambiguous but feasible solutions for the position of the center of the rectangle can be obtained by using the 6 TOA measurements from 4 anchors obtained by 3 antennas. As we can see in  Fig. \ref{fig:ambiguity}, the 4 solutions are so close to each other that it is difficult to identify the correct one without extra information. Therefore, we introduce measurements of multiple epochs to solve this problem, such that a unique solution can be determined without increasing the number of anchors.

To tackle the initialization and location ambiguity issue in the position determination problem, a new MEMA-TOA method, which estimates the parameter $ \bm{\theta}_{(k)}$ using the MEMA TOA along with inter-epoch constraints, is proposed in the next section.

 \begin{figure*}[hb]
	\hrulefill
	\newcounter{MYtempeqncnt}
	\setcounter{MYtempeqncnt}{\value{equation}}	
	\setcounter{equation}{10}
	\begin{align} \label{eq:J_KE}
		\mathtt{J}=&\underbrace{\sum_{k=1}^{K}\left(\Delta\bm{\rho}_{(k)}-\bm{g}_{(k)}\left(\bm{\theta}_{(k)}\right)\right)^{T} \bm{W}_{\mathrm{TOA}(k)}\left(\Delta\bm{\rho}_{(k)}-\bm{g}_{(k)}\left(\bm{\theta}_{(k)}\right)\right)}_{\mathtt{J}_\mathrm{TOA}}\\
		\nonumber
		&+ \underbrace{\sum_{k=2}^{K}\left(\Delta\tilde{\bm{\theta}}_{(k,k-1)}^{\mathrm{b}_{(k-1)}}-\bm{g}_{\mathrm{IP}}\left(\bm{\theta}_{(k)}, \bm{\theta}_{(k-1)}\right)\right)^{T} \bm{W}_{\mathrm{IP}(k)}\left(\Delta\tilde{\bm{\theta}}_{(k,k-1)}^{\mathrm{b}_{(k-1)}}-\bm{g}_{\mathrm{IP}}\left(\bm{\theta}_{(k)}, \bm{\theta}_{(k-1)}\right)\right)}_{\mathtt{J}_\mathrm{IP}}
	\end{align}
	\setcounter{equation}{\value{MYtempeqncnt}}	
\end{figure*}

\section{New MEMA-TOA Positioning Method}\label{method}

 \begin{figure}
  	\centering
  	\includegraphics[width=0.95\linewidth]{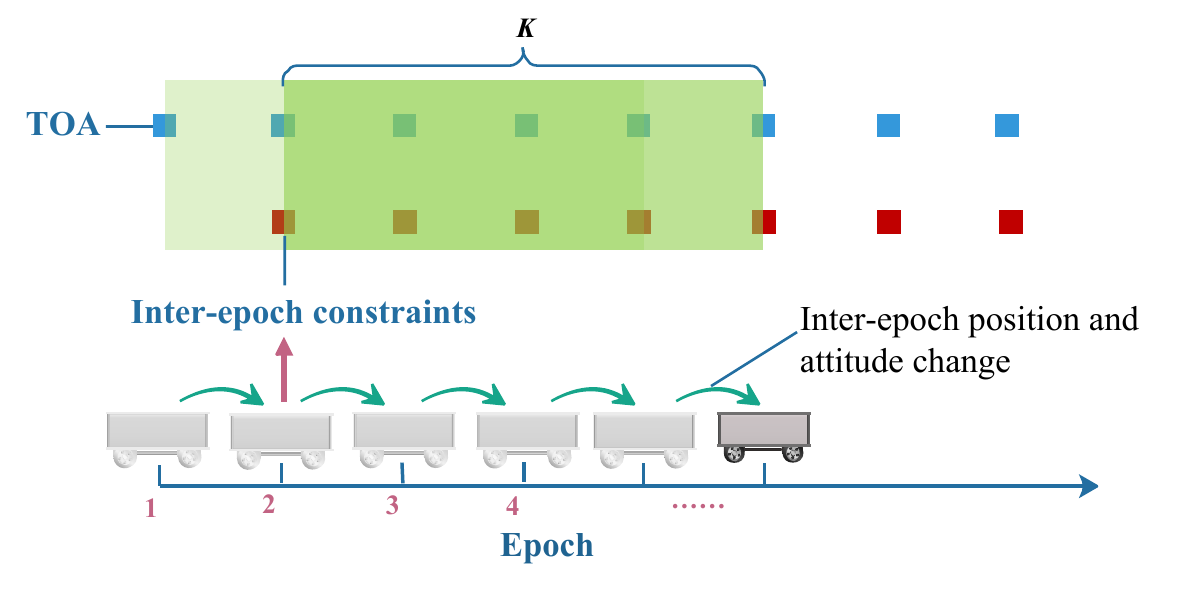}
  	\vspace{-0.2cm}
  	\caption{Measurements and constraints involved in the new MEMA-TOA method. The TOAs given by (\ref{eq:rhoij}) along with the inter-epoch constraints given by \eqref{eq:odo1} within the recent $K$ epochs are adopted for vehicle positioning.}
  	\label{fig:mhe}
  	\vspace{-0.3cm}
  \end{figure}

We develop a new positioning method, namely MEMA-TOA, based on multi-epoch and multi-antenna TOAs, in this section. We first derive the cost function of the positioning problem. Then, we develop the two steps of the MEMA-TOA method.

\subsection{Cost Function}

We first eliminate the influence of clock bias, by selecting one of the TOA measurements as reference for each epoch (e.g., the measurement between  the 1st antenna and its 1st visible anchor is chosen as the reference here, without loss of generality). We subtract \eqref{eq:rhoij} plugged with $i=1$ and $j=1$, from \eqref{eq:rhoij} with other $i$ and $j$, and come to the time difference of arrival (TDOA) measurement with respect to this reference as written by 
\begin{align}\label{eq:deltarhoij}
	\Delta\rho_{i1(k)}^{(j1)}&=\rho_{i(k)}^{(j)}-\rho_{1(k)}^{(1)}\nonumber\\
	&=\Vert\bm{p}_{(k)}^{(j)}-\bm{p}_{i(k)}\Vert-\Vert\bm{p}_{(k)}^{(1)}-\bm{p}_{1(k)}\Vert+\Delta\varepsilon^{(j1)}_{i1(k)}
	\text{,}
\end{align}
where $\Delta\rho_{i1(k)}^{(j1)}$ is the TDOA measurement and the noise term 
$\Delta\varepsilon^{(j1)}_{i1(k)}=\varepsilon^{(j)}_{i(k)}-\varepsilon^{(1)}_{1(k)}$.

All measurement equations of epoch $k$ as given by \eqref{eq:deltarhoij} are written in the collective form as
\begin{equation}\label{eq:deltarho_vecter}
	\Delta\bm{\rho}_{(k)}=\bm{g}_{(k)}\left(\bm{\theta}_{(k)}\right) +\Delta\bm{\varepsilon}_{(k)}
	\text{,}
\end{equation}
where
\begin{align}
    \Delta\bm{\rho}_{(k)}=&\left[ \Delta\rho_{11(k)}^{(21)},\ \cdots,\ \Delta\rho_{11(k)}^{\left(M_{1(k)}1\right)},\ \Delta\rho_{21(k)}^{(11)},\ \cdots,\ \right.\nonumber\\ 
    &\left.\Delta\rho_{21(k)}^{\left(M_{2(k)}1\right)},\ \cdots,\ \Delta\rho_{N1(k)}^{(11)},\ \cdots,\ \Delta\rho_{N1(k)}^{\left(M_{N(k)}1\right)}\right]^T\text{,}\nonumber
\end{align}
and $\bm{g}_{(k)}\left(\bm{\theta}_{(k)} \right)$ and $\Delta\bm{\varepsilon}_{(k)}$ are the corresponding vectors of the TDOA measurement function and noise at epoch $k$, respectively. The covariance matrix of $\Delta\bm{\varepsilon}_{(k)}$ is  
$\bm{Q}_{\Delta\bm{\varepsilon}_{(k)}}=\bm{E}_{(k)}\left( \sigma^2\bm{I}_{L_{(k)}+1}\right) \bm{E}_{(k)}^T$,
where
$\bm{E}_{(k)}=\left[ 
-\bm{1}_{{L}_{(k)}},\ \bm{I}_{{L}_{(k)}}\right] $, 
$L_{(k)}= \sum_{i=1}^N M_{i(k)}-1$ is the total number of TDOA measurements for $N$ antennas at epoch $ k $, $\boldsymbol{1}_{{L}_{(k)}}$ is an ${L}_{(k)}$-element column vector with all-one elements, and $\bm{I}_{{L}_{(k)}}$ is an ${{L}_{(k)}} \times{{L}_{(k)}}$ identity matrix.

In order to employ the inter-epoch position and attitude change constraint, we conduct the transformation from frame $\mathrm{n}$ to frame $\mathrm{b}$ for the vehicle position change as given by
\begin{align}\label{eq:deltapbk-1ton}
	\nonumber	\Delta\bm{p}_{(k,k-1)}^{\mathrm{b}_{(k-1)}}&=\bm{p}_{\mathrm{c}{(k)}}^{\mathrm{b}_{(k-1)}}-\bm{p}_{\mathrm{c}{(k-1)}}^{\mathrm{b}_{(k-1)}}\\
	&=\bm{R}_{(k-1)}^T\left(\bm{p}_{\mathrm{c}{(k)}}-\bm{p}_{\mathrm{c}{(k-1)}}\right)\text{. } 
\end{align}

By plugging \eqref{eq:deltapbk-1ton}, the inter-epoch constraints given by \eqref{eq:odo1} are then rewritten as
\begin{equation}\label{eq:odon}
	\begin{aligned}	
		\Delta\tilde{x}_{(k,k-1)}^{\mathrm{b}_{(k-1)}}=&\left[ \bm{R}_{(k-1)}^T\left(\bm{p}_{\mathrm{c}{(k)}}-\bm{p}_{\mathrm{c}{(k-1)}}\right)\right]_{1}+\varepsilon_{x_{(k)}} \text{, }\\ 	
		\Delta\tilde{y}_{(k,k-1)}^{\mathrm{b}_{(k-1)}}=&\left[ \bm{R}_{(k-1)}^T\left(\bm{p}_{\mathrm{c}{(k)}}-\bm{p}_{\mathrm{c}{(k-1)}}\right)\right] _{2}+\varepsilon_{y_{(k)}} \text{, }\\ 
		\Delta\tilde{\psi}_{(k,k-1)}^{\mathrm{b}_{(k-1)}}=& \psi_{(k)}-\psi_{(k-1)}+\varepsilon_{\psi_{(k)}} \text{. }
	\end{aligned}
\end{equation}

To simplify the expression, we rewrite \eqref{eq:odon} into the collective form as
\begin{align} \label{eq:rewrite_odo}
	\Delta\tilde{\bm{\theta}}_{(k,k-1)}^{\mathrm{b}_{(k-1)}}=\bm{g}_{\mathrm{IP}(k)}\left(\bm{\theta}_{(k)}, \bm{\theta}_{(k-1)} \right) +\bm{\varepsilon}_{\mathrm{IP}(k)}\text{,}
\end{align} 
where the subscript ``$\mathrm{IP}$'' stands for inter-epoch constraints, $\bm{g}_{\mathrm{IP(k)}} $ is a vector of the functions of $\bm{\theta}_{(k)}$ and $ \bm{\theta}_{(k-1)}$ as presented in (\ref{eq:odon}), 	
$\Delta\tilde{\bm{\theta}}_{(k,k-1)}^{\mathrm{b}_{(k-1)}}$ and $\bm{\varepsilon}_{\mathrm{IP}(k)}$ are the vectors of the inter-epoch constraints and the corresponding noise at epoch $k$, respectively, and the covariance matrix of $\bm{\varepsilon}_{\mathrm{IP}(k)}$ is  $\bm{Q}_{\bm{\varepsilon}_{\mathrm{IP}(k)}}= \operatorname{diag}\left( 	{\sigma _p^2},{\sigma _p^2},{\sigma _\psi ^2} 	 \right)$.

\begin{figure*}[hb]
	\setcounter{MYtempeqncnt}{\value{equation}}	
	\setcounter{equation}{15} 
	\hrulefill
		\begin{equation}
		\begin{aligned}\label{eq:h_G}
			&\bm{m}_{i(k)}=\\
			&\left\{ \begin{matrix}
			\left[\begin{matrix}
				\left(\Delta \rho_{11(k)}^{(2 1)}\right)^2-\bm{p}_{(k)}^{(2) T} \bm{p}_{(k)}^{(2)}+\bm{p}_{(k)}^{(1) T} \bm{p}_{(k)}^{(1)}-2 \left[ \Delta \bm{p}_{(k)}^{(2)}\right] _3 h-2\left(\bm{l}^{T} \otimes \bm{p}_{(k)}^{(1) T}-\bm{l}_{1}^{T} \otimes \bm{p}_{(k)}^{(2) T}\right) \boldsymbol{\alpha} \\
				\vdots \\
				\left(\Delta \rho_{11(k)}^{\left(M_{1(k)}1\right)}\right)^2-\bm{p}_{(k)}^{\left(M_{1(k)}\right) T} \bm{p}_{(k)}^{\left(M_{1(k)}\right)}+\bm{p}_{(k)}^{(1) T} \bm{p}_{(k)}^{(1)}-2 \left[ \Delta \bm{p}_{(k)}^{\left( M_{1(k)}\right) }\right] _3 h-2\left(\bm{l}_{1}^{T} \otimes \bm{p}_{(k)}^{(1) T}-\bm{l}_{1}^{T} \otimes \bm{p}_{(k)}^{\left(M_{l(k)}\right) T}\right) \boldsymbol{\alpha}
			\end{matrix}\right],  i=1 \text{,}	\\
		\left[\begin{matrix}
		\left(\Delta \rho_{i1(k)}^{(1 1)}\right)^2-\bm{p}_{(k)}^{(1) T} \bm{p}_{(k)}^{(1)}+\bm{p}_{(k)}^{(1) T} \bm{p}_{(k)}^{(1)}+\bm{l}_{1}^{T} \bm{l}_{1}-\bm{l}_{i}^{T} \bm{l}_{i}-2 \left[ \Delta \bm{p}_{(k)}^{(1)}\right] 
		_3 h-2\left(\bm{l}_{1}^{T} \otimes \bm{p}_{(k)}^{(1) T}-\bm{l}_{\bm{i}}^{T} \otimes \bm{p}_{(k)}^{(1) T}\right) \boldsymbol{\alpha} \\
			\vdots \\
		\left(\Delta \rho_{i1(k)}^{\left(M_{i(k)}1\right)}\right)^2-\bm{p}_{(k)}^{\left(M_{i}\right) T} \bm{p}_{(k)}^{\left(M_{i(k)}\right)}+\bm{p}_{(k)}^{(1) T} \bm{p}_{(k)}^{(1)}+\bm{l}_{1}^{T} \bm{l}_{1}-\bm{l}_{i}^{T} \bm{l}_{i}-2 \left[ \Delta \bm{p}_{(k)}^{\left(M_{i(k)}\right)}\right] _3h-2\left(\bm{l}_{1}^{T} \otimes \bm{p}_{(k)}^{(1) T}-\bm{l}_{i}^{T} \otimes \bm{p}_{(k)}^{\left(M_{i(k)}\right) T}\right) \boldsymbol{\alpha}
	\end{matrix}\right],\\
\qquad \qquad \qquad \qquad \qquad \qquad \qquad \qquad \qquad \qquad 
\qquad \qquad \qquad \qquad \qquad \qquad \qquad \qquad \qquad \qquad \qquad \qquad  i=2,\dots,N \text{,}\\
\end{matrix}	\right.\\
					&	\bm{G}_{i(k)}=\left\{ \begin{matrix} 2\left[\begin{matrix}
				{\left(\bm{l}_{1}^{T} \otimes \bm{p}_{(k)}^{(1) T}-\bm{l}_{1}^{T} \otimes \bm{p}_{(k)}^{(2) T}\right) \bm{\Gamma}} & \left[ \Delta \bm{p}_{(k)}^{(2)}\right]_{1: 2}^T & -\Delta \rho_{11(k)}^{(2 1)} & 0 \\
				\vdots & \vdots & \vdots & \vdots \\
				{\left(\bm{l}_{1}^{T} \otimes \bm{p}_{(k)}^{(1)T}-\bm{l}_{1}^{T} \otimes \bm{p}_{(k)}^{\left( M_{1(k)}\right)  T}\right) \bm{\Gamma}} & \left[ \Delta \bm{p}_{(k)}^{\left( M_{1(k)}\right) }\right] _{1: 2}^{T} & -\Delta \rho_{11(k)}^{\left(M_{1(k)} 1\right)} & 0
			\end{matrix}\right],  i=1 \text{,}\\
			2\left[\begin{matrix}
				{\left(\bm{l}_{1}^{T} \otimes \bm{p}_{(k)}^{(1) T}-\bm{l}_{i}^{T} \otimes \bm{p}_{(k)}^{(1) T}\right) \bm{\Gamma}} & \left[ \Delta \bm{p}_{(k)}^{(1)}\right]_{1: 2}^T& -\Delta \rho_{i1(k)}^{(1 1)} & \left(\bm{l}_{i}-\bm{l}_{1}\right)^{T} \\
				\vdots & \vdots & \vdots & \vdots \\
				{\left(\bm{l}_{1}^{T} \otimes \bm{p}_{(k)}^{(1) T}-\bm{l}_{i}^{T} \otimes \bm{p}_{(k)}^{\left(M_{i(k)}\right) T}\right)\bm{\Gamma}} & \left[ \Delta \bm{p}_{(k)}^{\left(M_{i(k)}\right)}\right] _{1: 2}^T & -\Delta \rho_{i1(k)}^{\left(M_{i(k)} 1\right)} & \left(\bm{l}_{i}-\bm{l}_{1}\right)^{T}
			\end{matrix}\right],i=2,\dots,N \text{.}
		\end{matrix}	\right.\\
		\end{aligned}
	\end{equation}
	\setcounter{equation}{\value{MYtempeqncnt}}	
\end{figure*}

As shown in Fig. \ref{fig:mhe}, the MEMA-TOA method utilizes TOAs from multiple recent epochs by introducing the inter-epoch constraints on the changes of position and attitude. The measurements of $K$ epochs and their inter-epoch constraints are collected, resembling a $K$-length sliding window moving forward one epoch at a time.

For $ K $ epochs, we denote the parameters to be estimated by
\begin{equation}\label{eq:thetaK}
\bm{\Theta}=\left[ 
\bm{\theta}_{(1)}^T ,\  \cdots,\ \bm{\theta}_{(K)}^T  
\right] ^T.
\end{equation} 

We then construct the MLE for the MEMA problem, utilizing the TOA measurements (\ref{eq:deltarho_vecter}) and inter-epoch constraints (\ref{eq:rewrite_odo}), as
\begin{equation}\label{eq:problem_KE}
		\hat{\bm{\Theta}}=\underset{\bm{\Theta}}{\arg \min } \mathtt{J}\text{,}
\end{equation}
where the cost function $\mathtt{J}$ for $ K $ epochs is given by (\ref{eq:J_KE}), in which $ \bm{W}_{\mathrm{TOA}(k)} $ and $ \bm{W}_{\mathrm{IP}(k)} $ are the weights for the TOA measurements and inter-epoch constraints, respectively, and both are determined by their noise covariance as $ \bm{W}_{\mathrm{TOA(k)}}=\bm{Q}_{\Delta\bm{\varepsilon}_{(k)}}^{-1} $  and  $ \bm{W}_{\mathrm{IP}(k)}= \bm{Q}_{\bm{\varepsilon}_{\mathrm{IP}(k)}}^{-1}$.

Note that the cost function $\mathtt{J}$ consists of two parts: $ \mathtt{J}_{\mathrm{TOA}}$ relating to the TOA measurements  and  $\mathtt{J}_{\mathrm{IP}} $ relating to the inter-epoch constraints, corresponding to the two summation terms in (\ref{eq:J_KE}).  Different  $ N $ and  $ K $ correspond to different cases.
When $  K=1 $ and $ N=1 $, $ \mathtt{J}_{\mathrm{IP}} $ vanishes, and the positioning problem then degenerates to the traditional case, which uses only the single-epoch and single-antenna measurements. This type of positioning problem is suitable for the application with sufficient measurements and no demand on attitude estimate, and has already been covered extensively in many researches such as \cite{kaplan2005GPS,Sahinoglu2011UWB,Shen2012TDOA}. When $  K=1 $ and $ N>1 $, $ \mathtt{J}_{\mathrm{IP}} $ also vanishes. It is a positioning problem based on single-epoch and multi-antenna TOA measurements, and was discussed in \cite{An2020DMA}. Finally, when $  K>1 $ and $ N>1 $,  the MEMA TOA measurements and inter-epoch constraints are used for positioning, and the corresponding problem is studied in this paper.


The proposed MEMA-TOA method is divided into two steps. Firstly, we develop an initialization method, namely MEMA-SDP, to obtain an initial guess. Secondly, we develop an iterative algorithm to refine the estimates based on the initial guess.

\subsection{Step 1: MEMA-SDP Initialization}

Due to the non-convexity of the original optimization problem in (\ref{eq:problem_KE}), finding the exactly optimal solution is challenging. In general, although an MLE is asymptotically efficient and does not depend on the initial guess, its iterative implementation for this positioning problem requires an accurate initial guess to avoid being trapped at a local minimum \cite{Kay1993SS, Zhao2021SDP, Jiang2019RBL_SDP2}.

SDP can achieve the global optimal solution of an approximate convex problem \cite{Wang2016SDP, Su2017SDP,Zhao2021SDP,Ke2020RBL_clock}, which is transformed from the original positioning problem by SDR and other approximations. To obtain a proper initialization, we transform and relax $ \mathtt{J}_{\mathrm{TOA}} $ and $ \mathtt{J}_{\mathrm{IP}} $ to form an SDP problem and then achieve the global optimal solution as an initial guess for the next step.

\subsubsection{Relaxation for TOA Cost Function}  \label{TOASDP}
We relax the non-convex cost function $\mathtt{J}_\mathrm{TOA}$ to a convex function with convex constraints by transformations and parameter substitutions.

For each TDOA measurement in \eqref{eq:J_KE}, we go back to its original form in (\ref{eq:deltarhoij}). Rewrite (\ref{eq:deltarhoij}) as	
	\setcounter{equation}{11} 
\begin{align}\label{eq:row1}
	\Delta \rho_{i 1(k)}^{(j 1)}+&\left\|\bm{p}^{(1)}_{(k)}-\left(\bm{R}_{(k)} \bm{l}_{1}+\bm{p}_{\mathrm{c}(k)}\right)\right\|\\
\nonumber &=\left\|\bm{p}^{(j)}_{(k)}-\left(\bm{R}_{(k)}  \bm{l}_{i}+\bm{p}_{\mathrm{c}(k)}\right)\right\|+\Delta\varepsilon_{i1(k)}^{(j1)}\text{.}
\end{align}

Squaring both sides of (\ref{eq:row1}), ignoring the second-order noise term and rearranging the equation, we have
	\begin{align}\label{eq:row3}
	\nonumber	&\left(\Delta \rho_{i 1(k)}^{(j 1)}\right)^{2}-\bm{p}^{(j) T}_{(k)} \bm{p}^{(j)}_{(k)}+\bm{p}^{(1) T}_{(k)} \bm{p}^{(1)}_{(k)}+\bm{l}_{1}^{T} \bm{l}_{1}-\bm{l}_{i}^{T} \bm{l}_{i} \\
\nonumber	=	&-2 \Delta \rho_{i 1(k)}^{(j 1)} r_{1(k)}^{(1)}+2\left(\bm{p}^{(1)}_{(k)}-\bm{p}^{(j)}_{(k)}\right)^{T} \bm{p}_{\mathrm{c}(k)}\\
&+2 \bm{p}^{(1) T}_{(k)} \bm{R}_{(k)} \bm{l}_{1}-2 \bm{p}^{(j) T}_{(k)} \bm{R}_{(k)} \bm{l}_{i}\\
\nonumber		&+2 \bm{p}_{\mathrm{c}(k)}^{T} \bm{R}_{(k)}\left(\bm{l}_{i}-\bm{l}_{1}\right)+2r_{i (k)}^{(j)}\Delta\varepsilon_{i 1(k)}^{(j 1)} \text{,}
	\end{align}
where ${r_{i(k)}^{(j)}} = \left\| \bm{p}^{\left( j \right )}_{(k)} - {\bm{p}_{i(k)}}\right\|$ is the distance between antenna $i$ and its $j$-th visible anchor at epoch $k$.

In order to convert (\ref{eq:row3}) to a linear and convex relation, we first vectorize $\bm{R}_{(k)} $ as $\operatorname{vec}\left(\bm{R}_{(k)}\right)=\boldsymbol{\alpha}+\bm{\Gamma u_{(k)}} $, in which 
 \begin{align}
 \nonumber &\bm{u}_{(k)}=\left[
	\mathrm{s}_{\psi_{(k)}},\ \mathrm{c}_{ \psi_{(k)}}	\right]^T\text{,}\\
\nonumber&{\bm{\Gamma }} = \left[ {\begin{matrix}
			0&{\mathrm{c}_ {\gamma} } &0 &
			{\mathrm{c}_ {\phi} }&{\mathrm{s}_{ \gamma} \mathrm{s}_ {\phi} }&0&{ - \mathrm{s}_ {\phi} }&{\mathrm{s}_{ \gamma} \mathrm{c}_ {\phi} }&0 \\ 
			{\mathrm{c}_ {\gamma} } & 0&0 & {\mathrm{s}_{ \gamma} \mathrm{s}_{ \phi} }&{ - \mathrm{c}_ {\phi }} & 	0& {\mathrm{s}_{ \gamma} \mathrm{c}_ {\phi} }
			&{\mathrm{s}_{ \phi} } &0
	\end{matrix}} \right]^T,\\
\nonumber &{\bm{\alpha }} = \left[ {
		0 ,\	0 ,\ {\mathrm{s}_ {\gamma} } ,\
		0 ,\	0 ,\		{ - \mathrm{c}_{ \gamma} \mathrm{s}_ {\phi }},\ 
		0 ,\ 0 ,\		{ - \mathrm{c}_{ \gamma} \mathrm{c}_{ \phi} }
} \right]^T.
\end{align}
More details on vectorization are presented in Appendix \ref{Appendix_VEC}.

Then, with the above vectorization of matrix $\bm{R}_{(k)} $, $\bm{p}_{(k)}^{(j) T} \bm{R}_{(k)} \bm{l}_{i}$ in \eqref{eq:row3} becomes \cite{Horn2012matrix}
\begin{equation}\label{eq:vecror_prl}
	\bm{p}_{(k)}^{(j) T} \bm{R}_{(k)} \bm{l}_{i}=\left(\bm{l}_{i}^{T} \otimes \bm{p}_{(k)}^{(j) T}\right) \cdot\operatorname{vec}\left(\bm{R}_{(k)}\right)\text{.}
\end{equation}

We then plug \eqref{eq:vecror_prl} into \eqref{eq:row3} and employ a new parameter
$$ \bm{f}_{(k)}=\left[
		\bm{u}_{(k)}^T ,\	x_{(k)},\ y_{(k)} ,\ r_{1(k)}^{(1 )},\ \bm{p}_{\mathrm{c}(k)}^{T} \bm{R}_{(k)}
\right]^{T}. $$
Thus, \eqref{eq:row3} becomes
\begin{equation}\label{eq:linear_TDOA}
	\bm{m}_{(k)}=\bm{G}_{(k)}\bm{ f}_{(k)}+\bm{B}_{(k)} \bm{E}_{(k)} \Delta \bm{\varepsilon}_{(k)}\text{,}
\end{equation}
where
\begin{equation}
\begin{aligned}
		\bm{m}_{(k)}=&\left[
		\bm{m}_{1(k)}^{T} ,\ \bm{m}_{2(k)}^{T} ,\ \cdots ,\ \bm{m}_{N(k)}^{T}
	\right]^{T} \text{,}\\
	 \bm{G}_{(k)}=&\left[
		\bm{G}_{1(k)}^{T} ,\ \bm{G}_{2(k)}^{T} ,\ \cdots ,\ \bm{G}_{N(k)}^{T}
	\right]^{T} \text{,}\\
		\bm{B}_{(k)} = &2\operatorname{diag}\left(
	r_{1(k)}^{(2)},\cdots,r_{1(k)}^{\left(M_{1(k)}\right)},r_{2(k)}^{(1)},\cdots,\right.\\
	&\left.r_{2(k)}^{\left(M_{2(k)}\right)},\cdots,
	r_{N(k)}^{(1)},\cdots, r_{N(k)}^{\left(M_{N(k)}\right)}\right) \text{,}\nonumber
\end{aligned}
\end{equation}
 $ \bm{m}_{i(k)}$ and $ \bm{G}_{i(k)} $, $ i=1,\cdots,N $ are given by (\ref{eq:h_G}), in which $ \Delta \bm{p}^{(j)}_{(k)} \triangleq \bm{p}^{(1)}_{(k)}-\bm{p}^{(j)}_{(k)} $, and $h$ is a constant denoting the known height.

Note that $ \bm{B}_{(k)} $ in \eqref{eq:linear_TDOA} indicates the contribution of each $\bm{m}_{(k)}$ in the cost function and contains the true distances $ r $ between antennas and their visible anchors. We are not able to know $ r $ when the position results of the antennas are not obtained yet. We can determine $ \bm{B}_{(k)} $  in the two practical cases, i.e., with and without clock bias knowledge. If there is prior knowledge on the clock bias, such as the clock bias estimate from the previous epoch, we can subtract the clock bias from the TOA to approximate $ r $. If there is no prior knowledge on clock bias, we have to approximate $ r $ using the TOA measurements directly. When the clock bias is large, the elements in $ \bm{B}_{(k)} $ are approximately the same, indicating equal weights for $\bm{m}_{(k)}$, as we will do in the numerical simulation in Section V. Although it may introduce some errors in the SDP results, the subsequent step will refine it iteratively.


At this stage, the approximated cost function relating to TOA measurements at epoch $ k $ becomes
	\setcounter{equation}{16} 
\begin{align}\label{eq:J_TDOA_original}
		\bar{\mathtt{J}}_{\mathrm{TOA}(k)}&=\\ \nonumber&\left(\bm{m}_{(k)}-\bm{G}_{(k)}\bm{f}_{(k)} \right)^T \bm { W }_{(k)} \left(\bm{m}_{(k)}-\bm{G}_{(k)}\bm{f}_{(k)} \right)\text{,}
\end{align}
where $\bm{W}_{(k)}=\bm{B}_{(k)}^{-1} \bm{Q}_{\Delta\varepsilon_{(k)}}^{-1} \bm{B}_{(k)}^{-1}$. It is convex, but is nonlinear in the parameters to be estimated, and thus not a standard form of SDP \cite{Boyd2004cvxbook}. Furthermore, we notice that $\bm{R}_{(k)}$, included in the parameter $\bm{f}_{(k)}$ to be optimized, has the non-convex constraints \cite{Farrell2008Aided}
\begin{align}\label{eq:problem_R}
	\bm{R}_{(k)}^{T} \bm{R}_{(k)}=\bm{I}_3\text{,}  \;
\operatorname{det}\left(\bm{R}_{(k)}\right)  =1 \text{.}
\end{align}

The TOA-related minimization problem becomes
	\begin{align}\label{eq:problem_TOA_SDP1}
		\mathop {\min }\limits_{{\bm{f}}_{\left( 1 \right)}, \ldots,{\bm{f}}_{\left( k \right)}} & \sum\limits_{k = 1}^K \bar{\mathtt{J}}_{\mathrm{TOA}(k)} 	\\
		\nonumber s.t. &\quad (\ref{eq:problem_R})\text{. }
	\end{align}
Note that (19) is non-convex. To obtain the standard-form SDP and transform the constraints to convex ones, we introduce a new parameter $\bm{F}_{(k)}=\bm{f}_{(k)}\bm{f}^{T}_{(k)}  $.

Utilizing $\bm{F}_{(k)}$ and the fact $\bm{x}^T\bm{W}\bm{x}=\operatorname{tr}\left(\bm{W}\bm{x}\bm{x}^T \right)$
for a vector $\bm{x}$, we then have 
\begin{align}
 \nonumber   &\left(\bm{m}_{(k)}-\bm{G}_{(k)}\bm{f}_{(k)} \right)^T \bm { W }_{(k)} \left(\bm{m}_{(k)}-\bm{G}_{(k)}\bm{f}_{(k)} \right)=\\
  \nonumber   &\operatorname{tr}\left\{\bm { W }_{(k)} \left(\bm{m}_{(k)}^{T}\bm{m}_{(k)}-2 \bm{G}_{(k)}\bm{f}_{(k)} \bm{m}_{(k)}^{T}
		+\bm{G}_{(k)}\bm{F}_{(k)} \bm{G}_{(k)}^{T}\right)\right\},
\end{align}
where $\operatorname{tr}\{\cdot\}$ is the trace of a matrix. We drop the constant term $\bm{m}_{(k)}^{T}\bm{m}_{(k)}$ to simplify the expression of the cost function as
\begin{align}\label{eq:J_TDOA}
		\bar{\mathtt{J}}_{\mathrm{TOA}(k)}&=\\
	\nonumber	&\operatorname{tr}\left\{\bm { W }_{(k)} \left(-2 \bm{G}_{(k)}\bm{f}_{(k)} \bm{m}_{(k)}^{T}
		+\bm{G}_{(k)}\bm{F}_{(k)} \bm{G}_{(k)}^{T}\right)\right\}\text{,}
\end{align}
where the parameters to be optimized are $ \bm{f}_{(k)}$ and $ \bm{F}_{(k)}$.

We then adopt the SDR method to relax the constraints in \eqref{eq:problem_R} and in $\bm{F}_{(k)}=\bm{f}_{(k)}\bm{f}^{T}_{(k)}  $ by  transforming the original ones and dropping the non-convex parts as follows \cite{Jiang2019RBL_SDP2,Ke2020RBL_clock}.


With the known  $\phi$ and $\gamma$  as well as $ 
	\bm{F}_{(k)}=\bm{f}_{(k)} \bm{f}^{T}_{(k)} $, the non-convex constraint $\bm{R}_{(k)}^{T} \bm{R}_{(k)}=\bm{I}_3$ in \eqref{eq:problem_R} is re-written as a convex one by 
\begin{equation}\label{eq:constrain1_TDOA}
	\operatorname{tr}\left(\left[ \bm{F}_{(k)}\right]_{1: 2,1: 2}\right)=1\text{.}
\end{equation}

Utilizing the fact that \cite{Horn2012matrix}
\begin{equation}\label{eq:Fff}
	\bm{F}_{(k)}=\bm{f}_{(k)} \bm{f}_{(k)}^{T}  \Leftrightarrow\left[
	\begin{matrix}
		\bm{F}_{(k)} & \bm{f}_{(k)} \\
		\bm{f}_{(k)}^{T} & 1
	\end{matrix}\right] \succeq \bm{0}_{9\times9}, \operatorname{rank}\left(\bm{F}_{(k)}\right)=1\text{,}
\end{equation}
the convex constraints for the TOA part at epoch $ k $ are then (\ref{eq:constrain1_TDOA}) and
\begin{align}\label{eq:constrain3_TDOA}
\left[\begin{matrix}
			\bm{F}_{(k)} & \bm{f}_{(k)} \\
			\bm{f}_{(k)}^{T} & 1
		\end{matrix}\right] \succeq \bm{0}_{9\times9} \text{.}
\end{align}
The non-convex constraints $ \operatorname{rank}\left( \bm{F}_{(k)}\right) =1 $ and $ \operatorname{det}\left( \bm{R}_{(k)}\right) =1 $ are dropped.

Furthermore, since $ r_{1(k)}^{(1)}=\left\|\bm{p}^{(1)}_{(k)}-\left(\bm{R}_{(k)} \bm{l}_{1}+\bm{p}_{\mathrm{c}(k)}\right)\right\| $, we have
	\begin{align}\label{eq:constrain2_TDOA}
	\left(r_{1(k)}^{(1)}\right)^2=& \quad[\bm{F}_{(k)}]_{5,5}\\
	\nonumber	 =&\quad \bm{p}^{(1) T}_{(k)} \bm{p}^{(1)}_{(k)}+\bm{l}_{1}^{T} \bm{l}_{1}+\operatorname{tr}\left\lbrace \left[ \bm{F}_{(k)}\right] _{3: 4,3: 4}\right\rbrace +h^{2}\\
		\nonumber	&-2\left[\bm{p}^{(1)}_{(k)}\right] _{1: 2}^{T} [\bm{f}_{(k)}]_{ 3: 4 }-2 \left[ \bm{p}^{(1) }_{(k)}\right]^T _3 h \\
		\nonumber	&-2\left(\bm{l}_{1}^{T} \otimes \bm{p}_{(k)}^{(1) T}\right)\left( \boldsymbol{\alpha}+\boldsymbol{\Gamma} \left[\bm{f}_{(k)}\right]_{1: 2} \right) \\
		\nonumber	&+2 \bm{l}_{1}^{T} \bm{R}_{(k)}^{T}\bm{p}_{\mathrm{c}(k)}\text{,}
	\end{align}
where 
$\bm{R}_{(k)}^{T}\bm{p}_{\mathrm{c}(k)}$  can be replaced by 
\begin{align}
	\nonumber\left[\bm{R}_{(k)}^{T}\bm{p}_{\mathrm{c}(k)}\right]_1=&
    \mathrm{c}_{\gamma }\left[ \bm{F}_{(k)}\right] _{2,3}+\mathrm{c}_{ \gamma}\left[ \bm{F}_{(k)}\right] _{1,4} +h \mathrm{s}_{ \gamma}, \\
\nonumber	\left[\bm{R}_{(k)}^{T}\bm{p}_{\mathrm{c}(k)}\right]_2=&\mathrm{c}_{ \phi} [\bm{F}_{(k)}]_{1,3}+\mathrm{s}_{ \gamma} \mathrm{s}_{\phi} [\bm{F}_{(k)}]_{2,3}-\mathrm{c}_{ \phi} [\bm{F}_{(k)}]_{2,4} \\
\nonumber	&+\mathrm{s}_{ \gamma} \mathrm{s}_{ \phi} [\bm{F}_{(k)}]_{1,4}-h \mathrm{c}_{ \gamma} \mathrm{s}_{ \phi },\\
\nonumber	\left[\bm{R}_{(k)}^{T}\bm{p}_{\mathrm{c}(k)}\right]_3=&	-\mathrm{s}_{ \phi} [\bm{F}_{(k)}]_{1,3}+\mathrm{s}_{ \gamma} \mathrm{c}_{ \phi} [\bm{F}_{(k)}]_{2,3}+\mathrm{s}_ {\phi} [\bm{F}_{(k)}]_{2,4}\\
 \nonumber &+\mathrm{s}_{ \gamma} \mathrm{c}_{ \phi} [\bm{F}_{(k)}]_{1,4}-h \mathrm{c}_{ \gamma} \mathrm{c}_ {\phi}.
\end{align}

Equation (\ref{eq:constrain2_TDOA}) provides a constraint on $\bm{f}_{(k)}$ and $\bm{F}_{(k)}$, and improves the estimation accuracy.


To sum up, for the TOA-related part in \eqref{eq:J_KE}, the optimazation problem becomes an SDP as
	\begin{align}\label{eq:problem_TOA_SDP_final}
		\mathop {\min }\limits_{\mathcal{X}_1} & \sum\limits_{k = 1}^K \bar{\mathtt{J}}_{\mathrm{TOA}(k)} 	\\
		\nonumber s.t. &\quad (\ref{eq:constrain1_TDOA})\text{, }(\ref{eq:constrain3_TDOA})\text{, and }(\ref{eq:constrain2_TDOA})\text{,}
	\end{align}
where $ \mathcal{X}_1=
		\left\{ {{{\bm{F}}_{\left( k \right)}},{{\bm{f}}_{\left( k \right)}}}  
		\right\},\; k = 1, \ldots,K$.
\subsubsection{Relaxation for Inter-epoch Cost Function}  
In this subsection, we convert the cost function $\mathtt{J_{\mathrm{IP}}} $ to a convex one and develop the corresponding constraints.

For each inter-epoch position constraints in \eqref{eq:J_KE}, we go back to the relationship in (\ref{eq:deltapbk-1ton}). Pre-multiplying both sides of (\ref{eq:deltapbk-1ton}) by $\bm{R}_{(k-1)}$ and utilizing the characteristic of rotation matrix that $\bm{R}_{(k-1)} \bm{R}_{(k-1)}^{T}=\bm{I}_3 $, it becomes
\begin{equation}\label{eq:deltapbk-1ton2}
	\bm{R}_{(k-1)}\Delta \tilde{\bm{p}}_{(k, k-1)}^{\mathrm{b}_{(k-1)}}=\bm{p}_{c(k)}-\bm{p}_{c(k-1)}\text{.}
\end{equation}

Utilizing the vectorization of $ \bm{R}_{(k-1)} $ and arranging the unknowns to the right side, we then have
\begin{align}\label{eq:ptodeltap}
	&\left(\left(\Delta \tilde{\bm{p}}_{(k, k-1)}^{\mathrm{b}_{(k-1)}} \right) ^T\otimes \bm{I}_3\right) \boldsymbol{\alpha}\\
\nonumber	&=	\bm{p}_{\mathrm{c}(k)}-\bm{p}_{c(k-1)}-\left(\left(\Delta \tilde{\bm{p}}_{(k, k-1)}^{\mathrm{b}_{(k-1)}}\right) ^T \otimes \bm{I}_3\right) \boldsymbol{\Gamma} \bm{u}_{(k-1)}\text{.}
\end{align}

For two-dimensional positioning, we take the noises of the inter-epoch position change into account and rearrange (\ref{eq:ptodeltap}) as
\begin{equation}\label{eq:linear_odo1}
\bm{m}_{p_{(k)}}=\bm{S}_{p_{(k)}} \bm{o}_{p_{(k)}}+\bm{\varepsilon}_{p(k)}\text{,}
\end{equation}
where
	\begin{align}
\nonumber	&\bm{S}_{(k)}=\left[\begin{matrix}
\bm{I}_2&-\left[\left( \Delta \tilde{\bm{p}}_{(k, k-1)}^{\mathrm{b}_{(k-1)}}\right) ^{T} \otimes \bm{I}_3\right]_{1:2,:} \boldsymbol{\Gamma} &-\bm{I}_2	
	\end{matrix}\right] \text{,}\\
\nonumber	&\bm{m}_{p_{(k)}}=\left[\left(\Delta \tilde{ \bm{p}}_{(k, k-1)}^{\mathrm{b}_{(k-1)}}\right) ^{T} \otimes \bm{I}_3\right]_{1:2,:} \boldsymbol{\alpha}\text{,} \\
\nonumber	&\bm{o}_{p_{(k)}}=\left[\begin{matrix}
		[\bm{f}_{(k)}]_{3: 4 }\\
		[\bm{f}_{(k-1)}]_{1: 2} \\
		[\bm{f}_{(k-1)}]_{3: 4}
	\end{matrix}\right]\text{,} 
\end{align}
and $ \bm{\varepsilon}_{p(k)} =\left[\varepsilon_{x(k)},\ \varepsilon_{y(k)}\right]^T$ is the noise of the position change constraint with 
the covariance matrix $\sigma _{p}^2\bm{I}_2$.

We then come to the yaw angle change constraints in \eqref{eq:J_KE}. According to the original form in (\ref{eq:odon}), the yaw angle change is rewritten as
\begin{equation}\label{eq:linear_odo2}
	m_{\psi_{(k)}}=\bm{S}_{\psi_{(k)}} \bm{o}_{\psi_{(k)}}+\varepsilon_{\psi_{(k)}}\text{,}
\end{equation}
where
	\begin{align}
	\nonumber	m_{\psi_{(k)}}&=\Delta \tilde{\psi}_{(k, k-1)}\text{,}\\
	\nonumber	\bm{S}_{\psi_{(k)}}&=\left[
		1 ,\ -1	\right]\text{,}\\
\nonumber\bm{o}_{\psi_{(k)}}&=\left[\psi_{(k)} ,\ \psi_{(k-1)}\right]^{T}\text{.}
\end{align}

We denote $ \bm{O}_{p_{(k)}} =\bm{o}_{p_{(k)}}\bm{o}_{p_{(k)}}^{T}  $ and $ \bm{O}_{\psi_{(k)}} = \bm{o}_{\psi_{(k)}} \bm{o}_{\psi_{(k)}}^{T}$. Utilizing (\ref{eq:linear_odo1}) and (\ref{eq:linear_odo2}), we can construct a function which is equivalent to the original cost function relating to the inter-epoch constraint at epoch $ k $  in \eqref{eq:J_KE}. Taking advantage of the fact that  $\bm{x}^T\bm{W}\bm{x}=\operatorname{tr}\left(\bm{W}\bm{x}\bm{x}^T \right)$ 
, the cost function becomes 
	\begin{align}\label{eq:J_odo}
	\nonumber 	&	\mathtt{J}_{\mathrm{IP}(k)}=\sigma^{-2} _{p}\operatorname{tr}\left(  -2{\bm{S}}_{p_{(k)}}{\bm{o}}_{p_{(k)}}\bm{m}_{p_{(k)}}^T + {{\bm{S}}_{p_{(k)}}}{{\bm{O}}_{p_{(k)}}}{\bm{S}}_{p_{(k)}}^T \right)  \\
&+\sigma _{\psi}^{-2} \operatorname{tr}{\left( {- 2{{\bm{S}}_{\psi_{(k)}}}{{\bm{o}}_{\psi_{(k)}}}{m}_{\psi_{(k)}}^T + {{\bm{S}}_{\psi_{(k)}}}{{\bm{O}}_{\psi_{(k)}}}{\bm{S}}_{\psi_{(k)}}^T} \right)} \text{.}
\end{align}
 The parameters to be optimized are $ {\bm{o}}_{p_{(k)}}$, $ {{\bm{O}}_{p_{(k)}}} $, $ {{\bm{o}}_{\psi_{( k )}}} $ and $ {{\bm{O}}_{\psi_{( k )}}} $.

Similar to the TOA constraint in (\ref{eq:Fff}), we obtain the positive semidefinite constraints as
 \begin{align}
 \label{eq:constrain5_TDOA}	&{\left[\begin{matrix}
 		\bm{O}_{p_{(k)}} & \bm{o}_{p_{(k)}} \\
 		\bm{o}_{p_{(k)}}^{T} & 1
 	\end{matrix}\right]\succeq \bm{0}_{7\times7} \quad k=2, \ldots, K}\text{,}\\
 \label{eq:constrain6_TDOA}	&{\left[\begin{matrix}
 		\bm{O}_{\psi_{(k)}} & \bm{o}_{\psi_{(k)}} \\
 		\bm{o}_{\psi_{(k)}}^{T} & 1
 	\end{matrix}\right] \succeq \bm{0}_{3\times3} \quad k=2, \ldots, K}\text{.}
 \end{align}

 Furthermore, according to the definition of  $ \bm{o}_{\psi_{(k)}} $, the constraint 
 \begin{equation}\label{eq:constrain4_TDOA}
 	\left[ \bm{o}_{\psi_{(k)}}\right] _2=\left[ \bm{o}_{\psi_{(k-1)}}\right] _1
 \end{equation}
 is added to improve the accuracy.
 
For the inter-epoch constraint part, the optimization problem becomes
\begin{align}\label{eq:problem_IP_SDP}
		\mathop {\min }\limits_{\mathcal{X}_2} & \sum\limits_{k = 2}^K \mathtt{J}_{\mathrm{IP}(k)} \\
		\nonumber s.t. &\quad		(\ref{eq:constrain5_TDOA})\text{, }
		(\ref{eq:constrain6_TDOA})\text{, and }(\ref{eq:constrain4_TDOA})\text{,}
	\end{align}
where $  \mathcal{X}_2=
		\left\{ {{{\bm{o}}_{p_{(k)}}},{{\bm{O}}_{p_{(k)}}},{{\bm{o}}_{\psi_{( k )}}},{{\bm{O}}_{\psi_{( k )}}}}  
		\right\},\; k = 2, \ldots,K$.
\subsubsection{Semidefinite Programming for MEMA problem}
Based on the above deduction, we combine the cost functions $ \bar{\mathtt{J}}_{\mathrm{TOA}} $ and $ \mathtt{J}_{\mathrm{IP}} $ as well as the corresponding constraints to construct a convex optimization problem as 
	\begin{align}\label{eq:problem_KE_SDP}
			\mathop {\min }\limits_{\mathcal{X}_1,\mathcal{X}_2} & \sum\limits_{k = 1}^K \bar{\mathtt{J}}_{\mathrm{TOA}(k)}   
		+ \sum\limits_{k = 2}^K \mathtt{J}_{\mathrm{IP}(k)} \\
		\nonumber s.t. &\quad (\ref{eq:constrain1_TDOA})\text{, }(\ref{eq:constrain3_TDOA})\text{, }(\ref{eq:constrain2_TDOA})\text{, }
		(\ref{eq:constrain5_TDOA})\text{, }
		(\ref{eq:constrain6_TDOA})\text{, and }(\ref{eq:constrain4_TDOA})\text{.}
	\end{align}

Once we solve the SDP problem, the global optimization of the problem (\ref{eq:problem_KE_SDP}) can be obtained. 

The problem (\ref{eq:problem_KE_SDP}) is not equivalent to the original problem (\ref{eq:problem_KE}) due to the relaxation and approximation, and there are errors between the SDP solution and the real solution. Therefore, we use it as the initialization and need to refine it to obtain the final positioning result.

\subsection{Step 2: Solution Refinement}
In this step, we refine the positioning results from the previous SDP step based on the multi-epoch TOA measurements and the inter-epoch constraints.

 For $ K $ epochs, the parameters to be estimated are $ \bm{\Theta}$ defined in \eqref{eq:thetaK}. The positioning problem for $ K $ epochs is the problem (\ref{eq:problem_KE}) with $ K>1 $ and $ N > 1 $.

The problem can be solved via iterative algorithms, such as the Gauss-Newton iterative method \cite{Bertsekas1999NLP}. The equations of TDOA measurements \eqref{eq:deltarhoij} and inter-epoch constraints \eqref{eq:odon} are linearized in the collective form as
\begin{equation}
	\delta \bm{z}= \bm{H}\cdot\delta\bm {\Theta} + {\bm{\varepsilon}}\text{,}
\end{equation}
where 
\begin{align}
\nonumber{\bm{H}} &= \left[  \bm{H}_{\mathrm{TOA}}^T,\ \bm{H}_{\mathrm{IP}}^T \right]^T \text{,}\\
\nonumber\delta \bm{z}&=\left[ 	\delta \bm{z}_{\mathrm{TOA}}^T,\ \delta \bm{z}_{\mathrm{IP}}^T \right]^T\text{,}
\end{align}	 
and the covariance matrix  $\bm{Q}_{\bm{\varepsilon}}$ is $ \operatorname{blkdiag}\left(
		\bm{Q}_{\bm{\varepsilon}_{\mathrm{TOA}} }, \bm {Q}_{ \bm{\varepsilon}_{\mathrm{IP}} } 
	 \right)$. 
$  \bm{H}_{\mathrm{TOA}}$, $\bm{H}_{\mathrm{IP}}$, $\delta \bm{z}_{\mathrm{TOA}}$, $\delta \bm{z}_{\mathrm{IP}}$, $\bm{Q}_{\bm{\varepsilon}_{\mathrm{TOA}} }$, $\bm {Q}_{\bm{\varepsilon}_{\mathrm{IP}} } $ as well as the details of linearization are derived in Appendix \ref{Appendix_ME}.

The cost function (\ref{eq:J_KE}) becomes
\begin{equation}\label{eq:problem_KE_ITER_LINEAR}
	\mathtt{J}_{\text{RE}} = \left( \delta \bm{z} - \bm{H}\cdot\delta \bm{\Theta} \right)^T\bm{Q}_{\bm{\varepsilon}}^{ - 1}\left( \delta \bm{z} -\bm{H}\cdot\delta \bm{\Theta} \right)\text{,}
\end{equation}
where the subscript ``RE'' stands for refinement.

The estimate of the increment is
\begin{equation}\label{eq:plobelm_KE_ITER_SOLUTION}
\delta \bm{\hat \Theta} = \left( \bm{H}^T\bm{Q}_{\bm{\varepsilon}}^{ - 1}\bm{H} \right)^{ - 1}\bm{H}^T\bm{Q}_{\bm{\varepsilon}}^{ - 1}\cdot\delta \bm{z}.
\end{equation}

Taking the estimated results of MEMA-SDP as the initial values and using (\ref{eq:plobelm_KE_ITER_SOLUTION}), the refined position and attitude can be obtained through multi-step iteration.

In reality, when there are limited computing resources but sufficient number of measurements in a single epoch, the iterative refinement step is reduced to the conventional method based on single-epoch TOA measurements, which has less computational complexity. However, for complex environments with insufficient measurements at a single epoch, the refinement with MEMA TOA measurements and inter-epoch constraints provides better accuracy and robustness at the cost of higher complexity.

Furthermore, it is worth mentioning that the proposed MEMA-TOA method conducts vehicle positioning, and solves the problem of location ambiguity caused by the insufficiency of TOAs at a single epoch. The high-precision positioning results can be obtained without a prior information of the initial value. Therefore, this new method can be used not only as a stand-alone positioning method but also as a position initialization for other positioning methods such as the EKF.

\section{Performance Analysis of MEMA-TOA Method}\label{performance}
In this section, we evaluate the performance of the proposed MEMA-TOA method by theoretically analyzing the accuracy and computational complexity.
\subsection{Accuracy}
We derive the Fisher information matrix for the new MEMA-TOA method. We compare it with the conventional single-epoch and multi-antenna TOA method (SEMA) \cite{An2020DMA} to show the superior positioning performance of the new method.
The attainable error variance, the CRLB and Fisher information matrix has a relation as \cite{Kay1993SS}
\begin{equation}\label{eq:CRLB}
\operatorname{var} \left( [\hat {\bm{\theta}}]_v  \right) \geqslant \mathrm{CRLB}\left( [\hat {\bm{\theta}}]_v \right) = {\left[ {{\mathtt{F}}^{ - 1}(\bm{\theta})} \right]_{v,v}},
\end{equation}
where $ \mathtt{F} $ is the Fisher information matrix, and the subscript ``$v$'' is the index.
The diagonal element of ${\mathtt{F}}^{ - 1} $  is the minimal variance that can be achieved theoretically in unbiased estimation.




When we estimate the position based on the TOA measurements of $ K $ epochs from $ N $ antennas along with the inter-epoch constraints, the Fisher information matrix is
	\begin{equation}\label{eq:FIM_ME}
		\mathtt{F}_{\mathrm{MEMA}} =\mathtt{F}_{\mathrm{MEMA,TOA}} + {\bm{H}}_{\mathrm{IP}}^T\bm{W}_{\mathrm{IP}}{{\bm{H}}_{\mathrm{IP}}},
	\end{equation}
where the subscript ``$\mathrm{MEMA} $'' stands for multi-epoch and multi-antenna, $\mathtt{F}_{\mathrm{MEMA,TOA}}$ is the Fisher information matrix derived by using only MEMA TOA measurements. Details of  $ 	\mathtt {F}_{\mathrm{MEMA}} $ are given in Appendix \ref{Appendix_CRLB}.

The diagonal element of ${\mathtt{F}}_{\mathrm{MEMA}}^{ - 1} $  is the minimal squared error that can be achieved in unbiased estimation for the MEMA problem. It is affected by the noises of TOA and inter-epoch position and attitude change constraints, the number of epochs and the number and geometry of visible anchors in each epoch. 

Furthermore, if there is no inter-epoch constraint,  
$\mathtt{F}_{\mathrm{MEMA}}=\mathtt{F}_{\mathrm{MEMA,TOA}} $ according to \eqref{eq:FIM_ME}.
The CRLB at epoch $ k $  is equal to the corresponding CRLB of the conventional case using single-epoch TOAs.

We also derive the Fisher information matrix for the conventional SEMA method \cite{An2020DMA} for comparison. For epoch $  k $, the parameters to be estimated are $\bm{\theta}_{(k)}= \left[
	x_{(k)},\ y_{(k)},\ \psi_{(k)}
\right]^T $. The Fisher information matrix is \cite{Kay1993SS}
	\begin{equation}\label{eq:FIM_SE}
		\mathtt{F}_{\mathrm{SEMA}( k )} = \bm{H}_{\left( k \right)}^T\bm{W}_{\mathrm{TOA}(k)}\bm{H}_{\left( k \right)}\text{,}
		\end{equation}
where the subscript ``$\mathrm{SEMA}$'' represents the conventional SEMA method, $ \bm{W}_{\mathrm{TOA}(k)} $ is the inverse of the covariance matrix of the TDOA measurement vector as defined in (\ref{eq:deltarho_vecter}). The details of $\bm{H}_{\left( k \right)}$ is shown in Appendix \ref{Appendix_ME}.

When MEMA TOA measurements and inter-epoch constraints are adopted, the CRLB of $ [{\bm{\Theta}}]_v $  has the relation
 \begin{equation}
	\mathrm{CRLB}_{{\mathrm{MEMA}},v} \leqslant \mathrm{CRLB}_{\mathrm{SEMA},v} \text{,}
\end{equation}
as derived in Appendix \ref{Appendix_CRLB}.

\textbf{Remark 1}: The theoretical estimation error of the new MEMA-TOA method is smaller that of the conventional SEMA method, showing the superior positioning accuracy of the new method.
\subsection{Computational Complexity}
We study the computational complexity of the proposed method. The complexity is shown in big $ O $ expressions with respect to the number of anchors $  M $, the number of antennas $ N $, the number of epochs  involved $ K $ and the localization dimension $ \eta=2 $ .  To investigate the worst case, it is assumed that all $N$ antennas can each receive all $M$ anchor signals, although this situation is unlikely to occur due to the obstacles.

In the first step, MEMA-SDP initialization, solving the SDP in (\ref{eq:problem_KE_SDP}) dominates the computation cost. We analyze the complexity of MEMA-SDP following the method in \cite{Jiang2019RBL_SDP2, Wang2016SDP} and \cite{ ben2001lcomplexity}. For each iteration in the inner-point algorithm utilized in SDP solving, the worst case complexity is about
\begin{equation}\label{eq:complexity_SDP}
\alpha^3+\alpha^2\sum\limits_{\iota=1}^{\xi}\beta_{\iota}^2+\alpha\sum\limits_{\iota=1}^{\xi}\beta_{\iota}^3,
\end{equation}
where $\alpha$ is the number of variables, $\xi$ is the number of constraints, and $\beta$ is the size of the constraint.  According to (\ref{eq:problem_KE_SDP}), in the proposed MEMA-SDP, the number of variables is $\alpha=(4\eta^2+18\eta+24)\cdot K-2\eta^2-7\eta-10$, where $K$ is the number of epochs involved. There are 6 types of constraints corresponding to (\ref{eq:constrain1_TDOA}),(\ref{eq:constrain3_TDOA}), (\ref{eq:constrain2_TDOA}),  (\ref{eq:constrain5_TDOA}), (\ref{eq:constrain6_TDOA}) and (\ref{eq:constrain4_TDOA}). The total number of constraints is $\xi=6K-3$. Calculating $\alpha$, $\beta_{\iota}$ and $\xi$ and substituting  them into (\ref{eq:complexity_SDP}), the complexity of one iteration is on the order of $O(\eta^6 K^3)$, and the iteration count is usually between 20 and 30 \cite{Biswas2006complexity}.

In the second step, the iterative algorithm is executed based on the results of Step 1 and the estimate of the increment (\ref{eq:plobelm_KE_ITER_SOLUTION}). The major computation for each iteration lies in the two inverse operations of the $\bm{Q}_\epsilon $ in (\ref{eq:plobelm_KE_ITER_SOLUTION}), which is $O(2\varpi^3)$\cite{quintana2001matrix_inversion}, where $\varpi= K(MN-1)+3(K-1) $.

\section{Numerical Simulations}\label{simulation}
We first design an extreme scene, in which there are only four anchors, and the number of visible anchors at each epoch is very limited. In this scene, we verify that our method successfully removes the ambiguous locations and achieves the theoretical accuracy. We then design a more practical scene, which simulates an unmanned cargo port. In this scene, we plan a path for the vehicle to be located and simulate the visible anchors at each epoch according to the vehicle position, attitude and the obstacles in the environment to test the performance of our method.

To the best of the authors’ knowledge, there is no other method in the literature, utilizing MEMA TOAs and inter-epoch constraints to deal with the vehicle positioning problem under the dense obstacle environment. Therefore, we compare the new MEMA-TOA method with the conventional SEMA method \cite{An2020DMA} to show the outstanding performance of our proposed method.

\subsection{Positioning in an Extreme Case with Minimum Number of Anchors }\label{simple_scene}
\subsubsection{Scene Setting}
\begin{figure}
	\centering
	\includegraphics[width=0.99\linewidth]{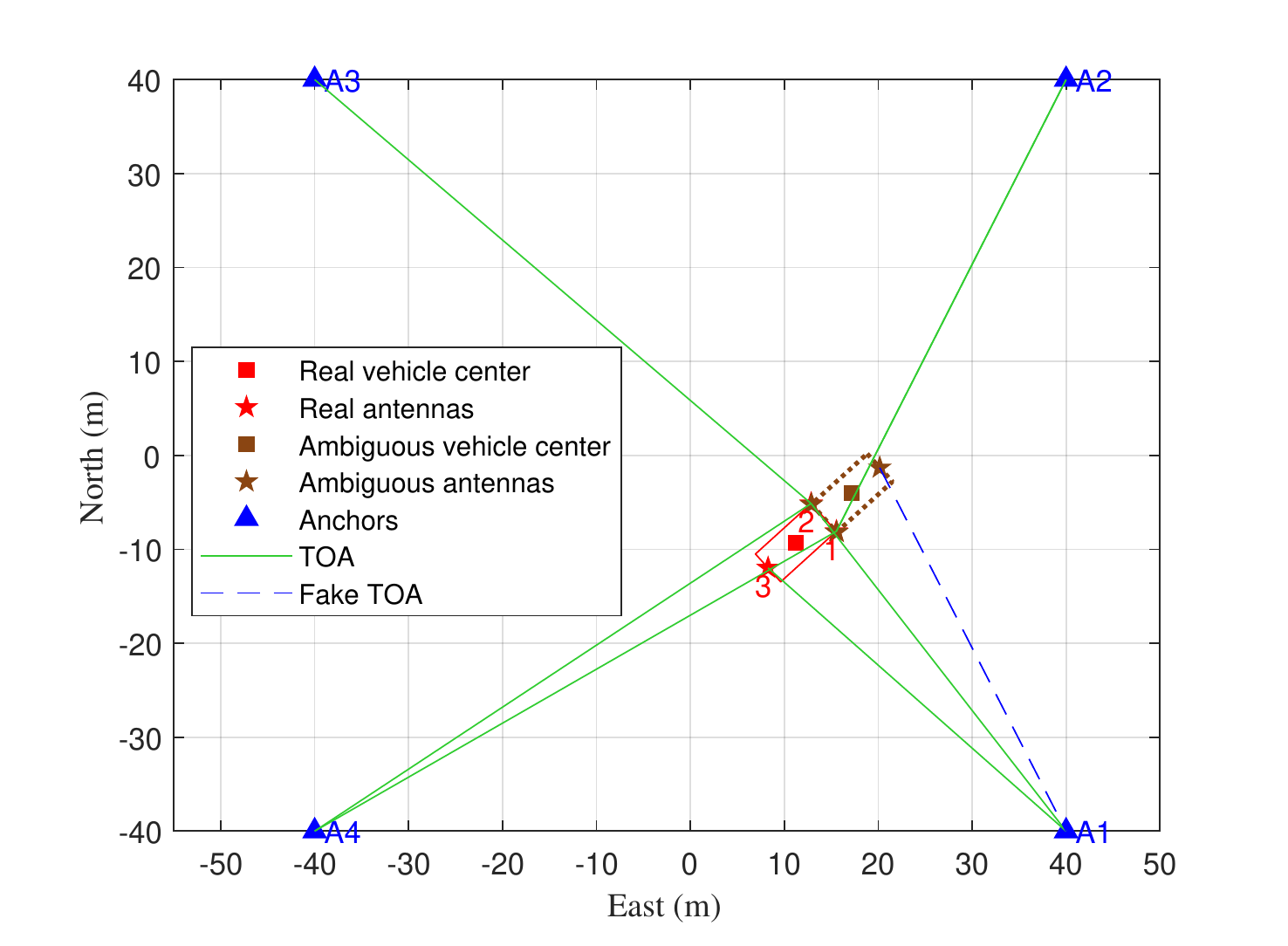}
	\vspace{-0.8 cm}
	\caption{A typical 2D vehicle positioning scene with four anchors.
	 The ambiguous location appears in the conventional SEMA method since the fake TOA (blue-dashed line) may be mistaken as the true TOA measurement between A1 and antenna 3.
	}
	\label{fig:scene_fouranchor}
	\vspace{-0.3 cm}
\end{figure}
We setup an extreme scene as shown in Fig. \ref{fig:scene_fouranchor}, in which there are only 4 anchors forming an area of 80 m $\times$ 80 m. The anchors have known positions and are synchronized to a common clock source. The center of the vehicle body is the reference point and the origin of the body frame. 
The coordinates of the three on-board antennas in  frame  $\mathrm{b}$  are  $[
    4, -2]^T$  m,  $[4, 2]^T$ m and $[-4,0]^T$ m,  respectively. The coordinates of the four anchors  in  frame  $\mathrm{n}$  are  $[40,-40]^T$ m,  $[40, 40]^T$ m,  $[-40,40]^T$  m and  $[-40,-40]^T$ m, respectively. The clock bias between the receiver and the anchors is set to 149.90 m. Both the TOA measurements and the inter-epoch change constraints are updated at a 1-Hz rate. 

We simulate the case that antennas can not receive the signals from all the anchors. As shown in Fig. \ref{fig:scene_fouranchor}, there are only 6 TOAs for this epoch. And the antennas at the ambiguous location illustrated by the brown squares in the figure can receive almost the same set of TOAs (the fake TOA is mistaken as the true TOA since it is almost identical with the true TOA between A1 and antenna 3 in the figure). Thus, utilizing the TOAs of this single epoch, the position results may fall into multiple locations as shown by the ambiguous location in the figure. 
\begin{table}[ht]
	\centering
	\begin{threeparttable}	
		\caption{Simulated positions and yaw angles of the vehicle and visible anchors at each epoch.}
		\label{table:scene1}
		\centering
		\vspace{0.2cm}
	\begin{tabular}{p{0.5cm} c c l}
			\toprule
			Epoch & Position (m) & Yaw (rad) & Visible anchors
			 \\
	\midrule 1& $[11.24,-9.29]^T$&0.74&$\{2,4\},\{1,3,4\},\{1\}$\\
	2& $[12.46,-9.05]^T$&0.65&$\{2,4\},\{3\},\{1,2,3,4\}$\\
	3& $[13.78,-9.07]^T$&0.54&$\{1,2,3,4\},\{ 1,2,3\},\{2,3,4\}$\\
4& $[15.12,-9.26]^T$&0.41&$\{1,4\},\{3\},\{2,3,4\}$\\
\bottomrule
		\end{tabular}
	\begin{tablenotes}[para,flushleft]
	Note: The numbers in the \{\} in the last column are the visible anchor numbers for each antenna. 
\end{tablenotes}	
		\end{threeparttable}
	\vspace{-0.2cm}
\end{table}

We set 4 epochs for this scene, the details of the simulated positions and yaw angles of the vehicle, and the visible anchors at each epoch are shown in Table \ref{table:scene1}.
Based on the above configuration, we conduct 500 Monte Carlo simulations. 

\subsubsection{Simulation Results}\label{results_A}
\begin{figure}
	\centering
	\subfloat[ MEMA-SDP with 2 epochs]{
	\includegraphics[width=0.46\linewidth]{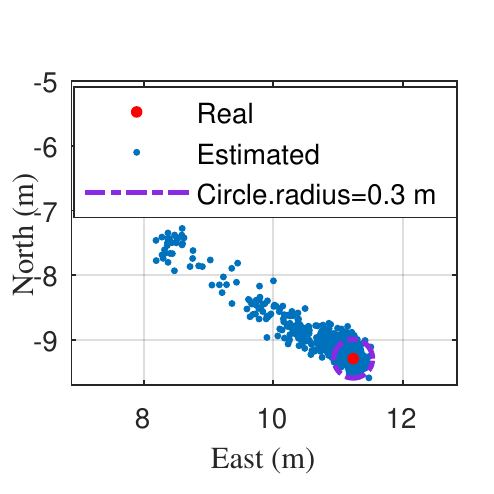}\label{fig:2E_SDP}}
		\hspace{0.02\linewidth}
	\subfloat[MEMA-SDP with 3 epochs]{
	\includegraphics[width=0.46\linewidth]{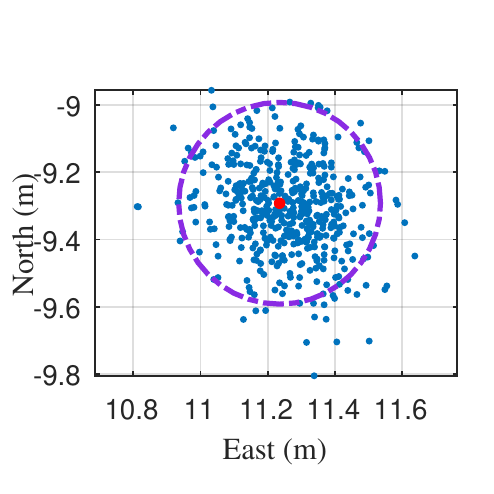}\label{fig:3E_SDP}}
		\hspace{0.02\linewidth}
	\subfloat[MEMA-SDP with 4 epochs]{
	\includegraphics[width=0.46\linewidth]{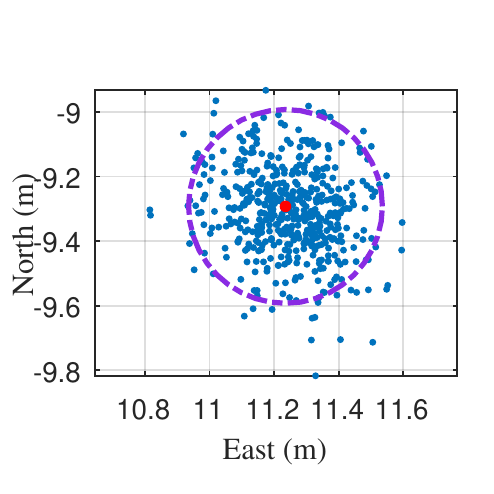}\label{fig:4E_SDP}}
	\hspace{0.02\linewidth}
	\subfloat[Conventional SEMA]{
	\includegraphics[width=0.46\linewidth]{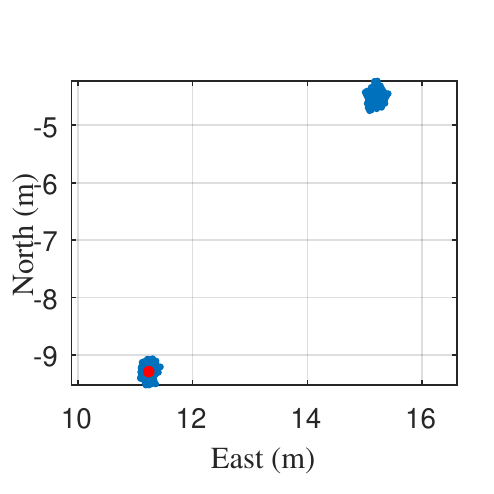}\label{fig:two_locations}}
		\vspace{0cm}
	\caption{Position results of the new MEMA-SDP (with TOAs from different numbers of epochs) and the conventional SEMA.
	(a) MEMA-SDP with 2 epochs: Over $52\%$ of the position estimates are outside the 0.3 m circle centered at the real location. (b) MEMA-SDP with 3 epochs: $90.6\%$  of the position results are inside the circle. (c) MEMA-SDP with 4 epochs: $91.2\%$ of the position results are inside the circle. (d) Conventional SEMA: The two separate groups represent the correct and the ambiguous locations, respectively. The new MEMA-SDP has higher position accuracy with more epochs of TOAs, and outperforms the conventional SEMA. }
	\label{fig:SDP}
	\vspace{-0.3cm}
\end{figure}

We first apply our MEMA-SDP method to obtain the initial position guess, and the results with different numbers of epochs, i.e., $K=2,3$ and $4$ are presented in Fig. \ref{fig:SDP} \subref{fig:2E_SDP}, \subref{fig:3E_SDP} and  \subref{fig:4E_SDP}, respectively. The standard deviations of the noises for the TOA measurement, the inter-epoch position and attitude constraints are set to $\sigma= 0.1$ m, $\sigma_p= 0.1$ m and $\sigma_{\psi}= 0.1 $ rad, respectively. As illustrated in Fig. \ref{fig:SDP} \subref{fig:2E_SDP}, the results of MEMA-SDP with 2 epochs are not all close to the real location, and more than $52\%$ of the estimates are more than 0.3 m away from the real location. Both the initial guess results from MEMA-SDP with 3 epochs and 4 epochs are closer to the real location than those from the case with 2 epochs, as shown in Fig. \ref{fig:SDP} \subref{fig:3E_SDP} and \subref{fig:4E_SDP}. About $90.6\%$ and $91.2\%$ of the estimates are inside the circle, respectively. This result shows that with more measurements from more epochs, the MEMA-SDP obtains a more accurate initial position.
For comparison, we depict the results of the conventional SEMA method \cite{An2020DMA} in Fig. \ref{fig:SDP} \subref{fig:two_locations}. The SEMA is initialized with a random location, which is drawn from uniformly distributed random coordinates in the area formed by the anchors. We can see that the results cluster into two groups. One group is around the real location, and the other group far apart indicates the ambiguous solution. It shows that the new MEMA-SDP can effectively remove the location ambiguity, and obtains an initial position close to the real location, compared with the conventional SEMA method.


\begin{figure}
	\centering
	\includegraphics[width=0.99\linewidth]{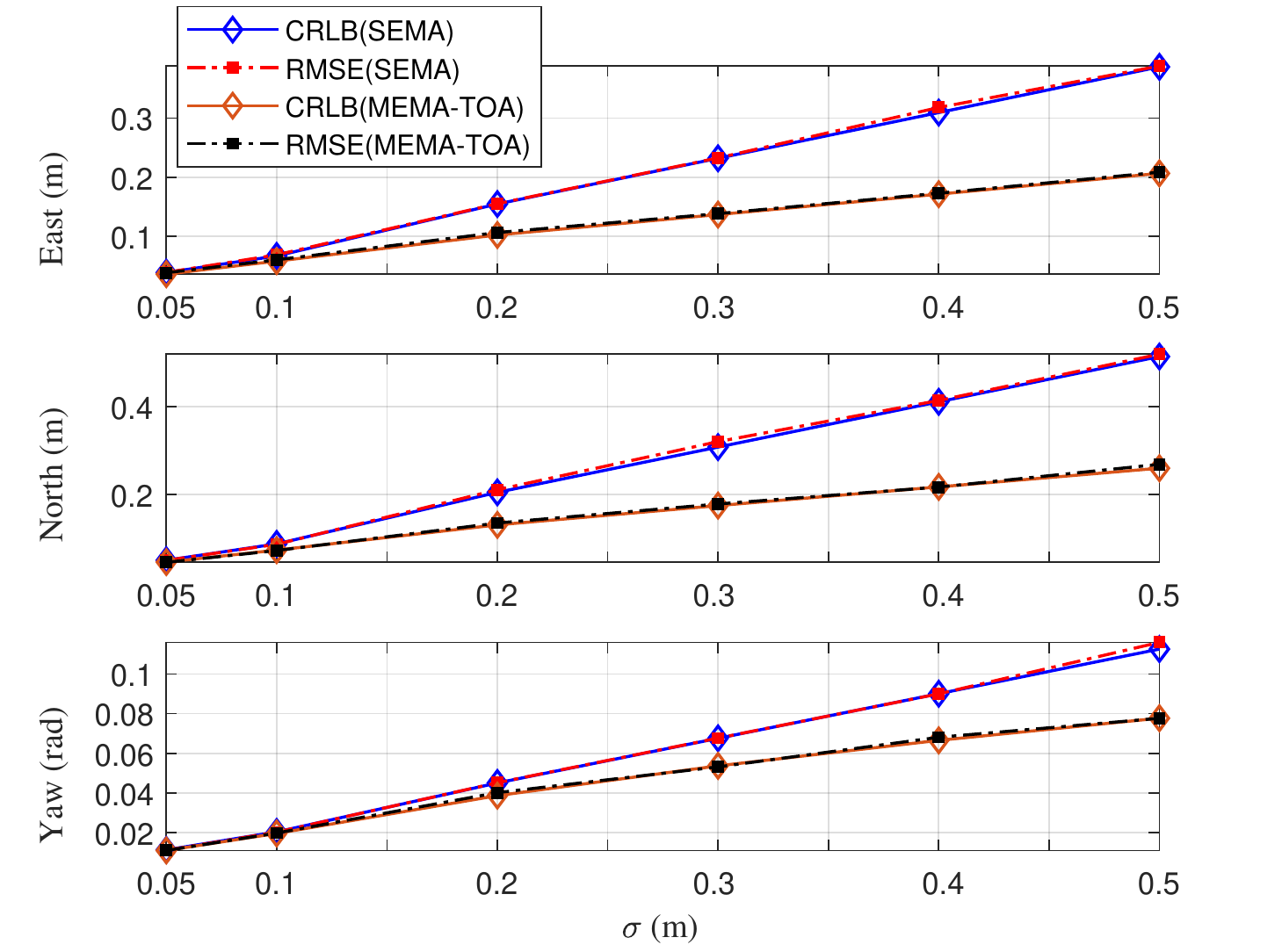}
	\vspace{-0.8cm}
	\caption{Position and yaw RMSEs vs. TOA measurement noise. Results are from 500 Monte-Carlo simulations. The position and yaw estimation accuracies of the new MEMA-TOA method reach the CRLBs, and are higher than those of the conventional SEMA. }
	\label{fig:CRLB}
	\vspace{0cm}
\end{figure}

After initialization by the MEMA-SDP, we apply the proposed refinement step to the initial position. The positioning root‐mean‐square error (RMSE) of the proposed MEMA-TOA method from 500 Monte Carlo simulation tests are calculated and compared to the CRLB computed based on (\ref{eq:CRLB}), (\ref{eq:FIM_ME}) and (\ref{eq:FIM_SE}).  

The positioning RMSEs and the theoretical CRLBs with different standard deviation of TOA noise are shown in Fig. \ref{fig:CRLB} with the inter-epoch constraint noise $\sigma_p= 0.1$ m and $\sigma_{\psi}= 0.1 $ rad, respectively. It can be seen from the figure that our MEMA-TOA method can achieve the theoretical CRLB. We show that the RMSE of the new MEMA-TOA is lower that that of SEMA. It verifies that the estimation error of the new MEMA-TOA method is lower that of the conventional SEMA method, consistent with the analysis in Section \ref{performance}.

\begin{figure}
	\centering
	\includegraphics[width=0.99\linewidth]{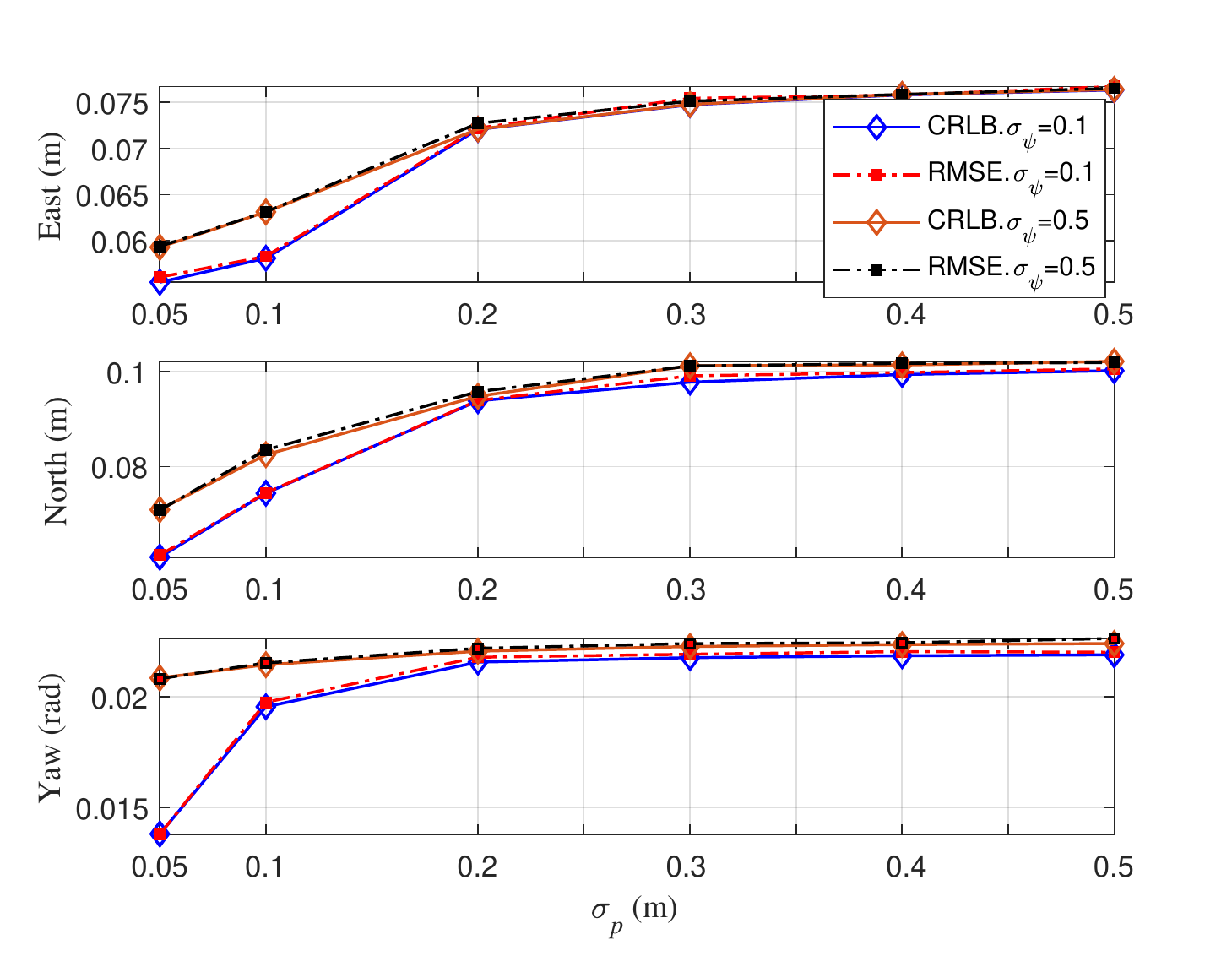}
	\vspace{-0.8cm}
	\caption{Position and yaw RMSE results from the MEMA-TOA method with different noise of the inter-epoch constraints. The inter-epoch position change noise $\sigma_p$ varies from 0.05 m to 0.5 m, and the yaw change noise $\sigma_{\psi}$ is set to 0.1 rad and 0.5 rad. The estimation accuracy in different cases reaches the CRLB, and decreases with the increasing noise of the inter-epoch constraint. }
	\label{fig:CRLB_var}
	\vspace{0cm}
\end{figure}

We further investigate the positioning performance of the new MEMA-TOA method with different noises of inter-epoch constraints. We fix the TOA noise as $\sigma=0.1$ m and vary the standard deviation of inter-epoch position and attitude change noises. The position and yaw estimation RMSEs are shown in Fig. \ref{fig:CRLB_var}. 
As illustrated in the figure, the estimation accuracy decreases with the increase of the inter-epoch constraint noise. With the same inter-epoch position noise $\sigma_p$, lager inter-epoch yaw noise $\sigma_{\psi}$ leads to larger estimation error. The estimation accuracy reaches the CRLB, showing the optimality of the new method. 

 All the simulations are implemented using Matlab 2017a on a personal computer with a 2.5-GHz i7-6500U CPU and 8GB RAM. The MEMA-SDP step is realized using the Matlab toolbox CVX \cite{Boyd2014cvx} with the solver SeDuMi \cite{Sturm1999solver} with default precision. We record the computation time of 500 simulation runs for our MEMA-TOA method. The average run time is 987.0 ms, and the number of iterations in Step 2 of MEMA-TOA is about 8. The average run time of the conventional SEMA method is 1.3 ms, which is lower than that of the MEMA-TOA method, since the SEMA executes only the iterative algorithm based on the single-epoch and multi-antenna TOAs. Also note that the CVX we use to solve the SDP is a universal solver, which is not specially designed for an efficient solution to this specific problem.

\subsection{Positioning in a Practical Harsh Environment:  Unmanned Cargo Port}\label{simulatin_port}
\subsubsection{Scene Setting} 
\begin{figure}
	\centering
	\includegraphics[width=0.99\linewidth]{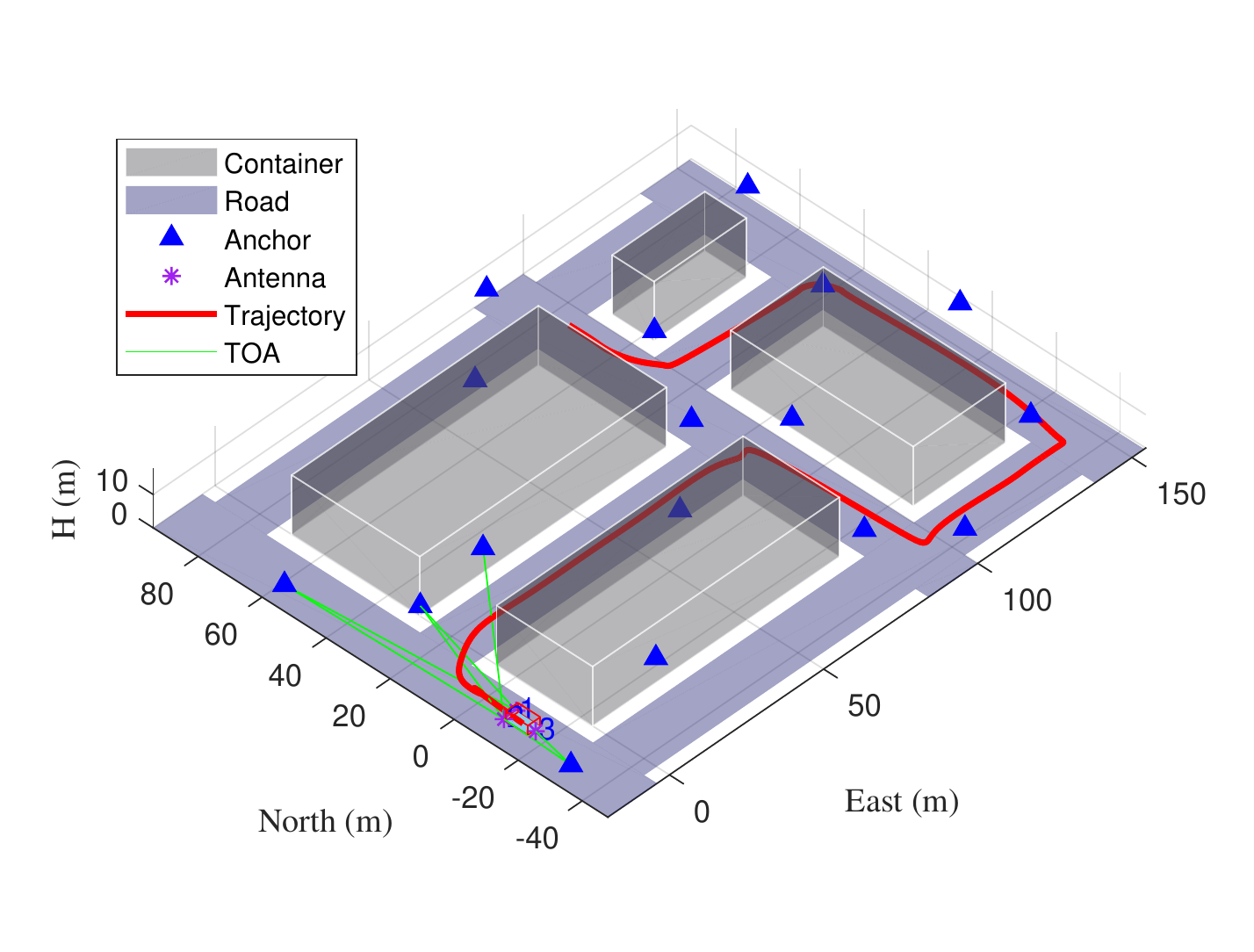}
	\vspace{-0.5cm}
	\caption{Simulation scene for an unmanned cargo port.
	Due to the obstacles in the environment and on the vehicle, the fixed antennas on the vehicle can only receive the signals from a few anchors.}
	\label{fig:scene_port}
	\vspace{-0.3cm}
\end{figure}

As shown in Fig. \ref{fig:scene_port}, we construct a simulation scene to simulate an unmanned cargo port. There are several containers represented by the gray cuboids stacked on the port. The vehicle to be located is moving along the roads between the containers.
Considering the limitation of gantry cranes in the port environment, the heights of anchors are limited to a certain extent. Consequently, the roadside containers and goods on the vehicle will block the signals from the anchors. 
Three synchronized receiving antennas are mounted on the vehicle to receive more signals. 

The vehicle is 8 m long, 4 m wide and 2 m high. It moves on the 16 m wide roadway. The goods with 7 m in length, 3 m in width and 2.8 m in height are loaded on the vehicle. Three antennas are installed on the plate of the vehicle. The coordinates of the three antennas in frame $\mathrm{b}$ are $[4,2,-0.15]^T$ m, $[4,-2,-0.15]^T$ m and$ [-4,0,-0.15]^T$ m, respectively. Considering the height limitation of gantry cranes in the port environment, we set the height of the anchors to 8 m, and the east and north coordinates of the anchors in frame $\mathrm{n}$ are presented in Table \ref{table_anchor_scene_3A} with the unit of meter. 

 \begin{table}[ht]
 	\centering
 	\begin{threeparttable}	
 		\caption{Coordinates of anchors in the simulated scene of unmanned cargo port (frame $ \mathrm{n} $).}
 		\label{table_anchor_scene_3A}
 		\centering
 		\vspace{-0.3cm}
 		\begin{tabular}{c  r r  c r r }
 			\toprule
 			{ Anchor No. } & {East }&{North}&{ Anchor No. } & {East }&{North }\\
 			\hline
 			1 & 150& 72&10& 100&53\\
 			2 & -4& 26&11&84&26\\
 			3 & -19.5&-36&12 & 84&-28\\
 			4&60 &  70.5&13 & 100 &  10\\
 			5& 154.5& 10&14& 20 &  29.5\\
 			6 & 137.5&36.5&15&-20& 53\\
 			7 &  138& -28&16 & 20&-24.5\\
 			8&  100& -44&17 &  60& 6.5\\
 			9&  84&90& & & \\
 			 			\bottomrule
 		\end{tabular}	
 	\end{threeparttable}
 	\vspace{-0.2cm}
 \end{table}
 
\begin{figure}
 	\centering
 	\includegraphics[width=0.99\linewidth]{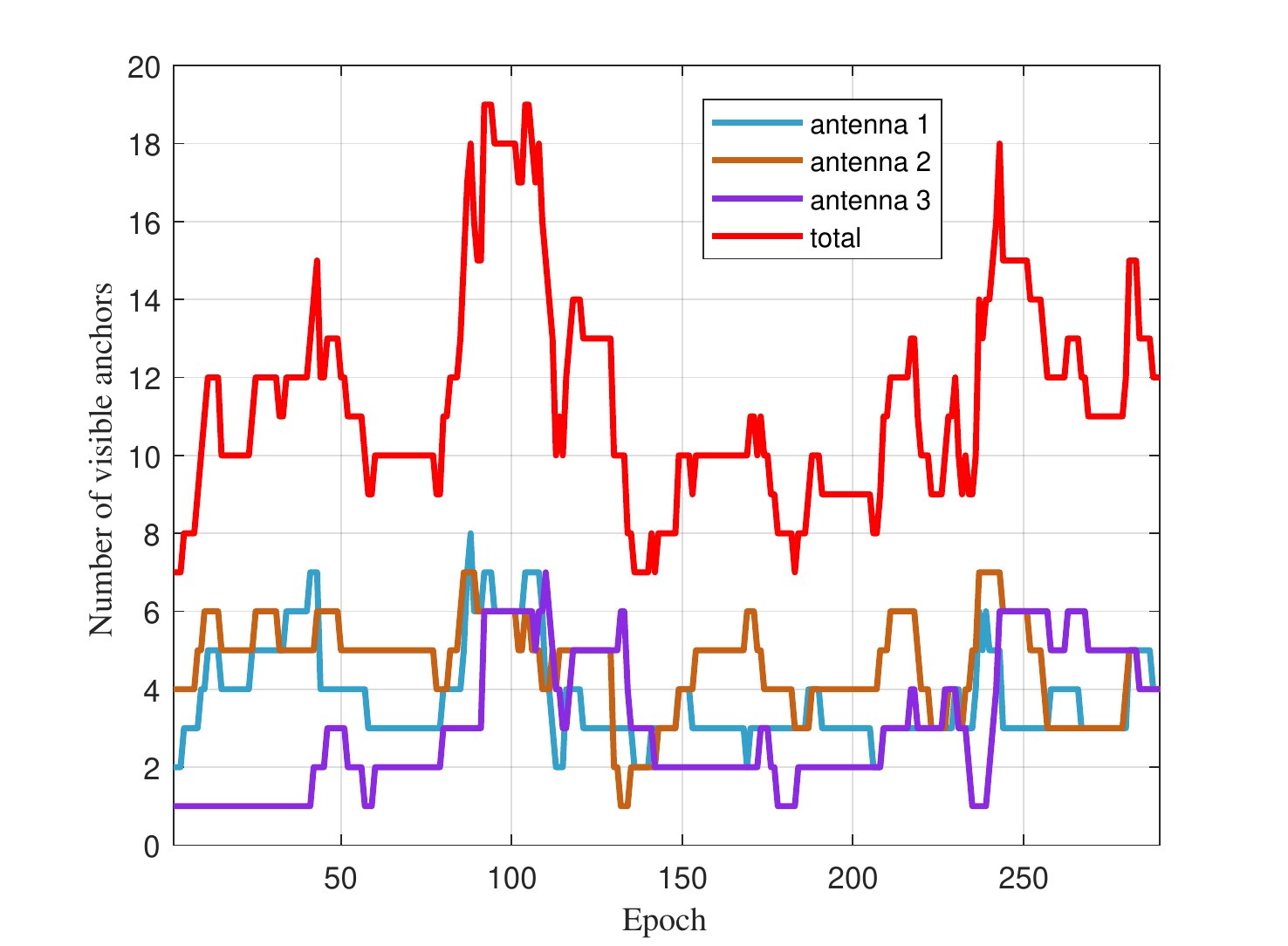}
 	\vspace{-0.5 cm}
 	\caption{Number of visible anchors at different antennas in the unmanned port simulation scene. Position and attitude results cannot be obtained using only a single antenna, since the number of visible anchors for a single antenna often falls below 4.}
 	\label{fig:number}
 	\vspace{-0.3cm}
 \end{figure}

In this simulation scene, we set $M=17$ and $N=3$  for problem (\ref{eq:problem_KE}). As shown by the red line in Fig. \ref{fig:scene_port}, the vehicle moves along a trajectory on the road for 290 s, i.e., the number of total epochs is 290. The clock bias between the receiver and the anchors is set to 149.90 m. Both the TOA measurements and the inter-epoch change constraints are updated at a 1-Hz rate. The standard deviations of the noises for the TOA measurement, the inter-epoch position and attitude constraints are set to $\sigma= 0.1$ m, $\sigma_p= 0.1$ m and $\sigma_{\psi}= 0.1 $ rad, respectively. The measurements for each antenna at each epoch are generated according to the the specifications of the commercial off-the-shelf UWB chip, IMU and the characteristics of the real-world implemented system based on it \cite{Chen2020odo,DW2020UWB,pala2020UWBaccurate,Sidorenko2020UWBTOA}. Due to the change of the relative position between the vehicle and the anchors and containers, each epoch has a different geometry of the anchors and represents different situations. The number $\mathcal{M}_{i(k)}$ of visible anchors for antenna $i$ at epoch $k$ varies with the motion of the vehicle, as shown in Fig. \ref{fig:number}.  We can see that with multiple antennas, the total number of visible anchors is increased compared with the single-antenna cases.


\subsubsection{Simulation Results} 
\begin{figure}
	\centering
	\includegraphics[width=0.99\linewidth]{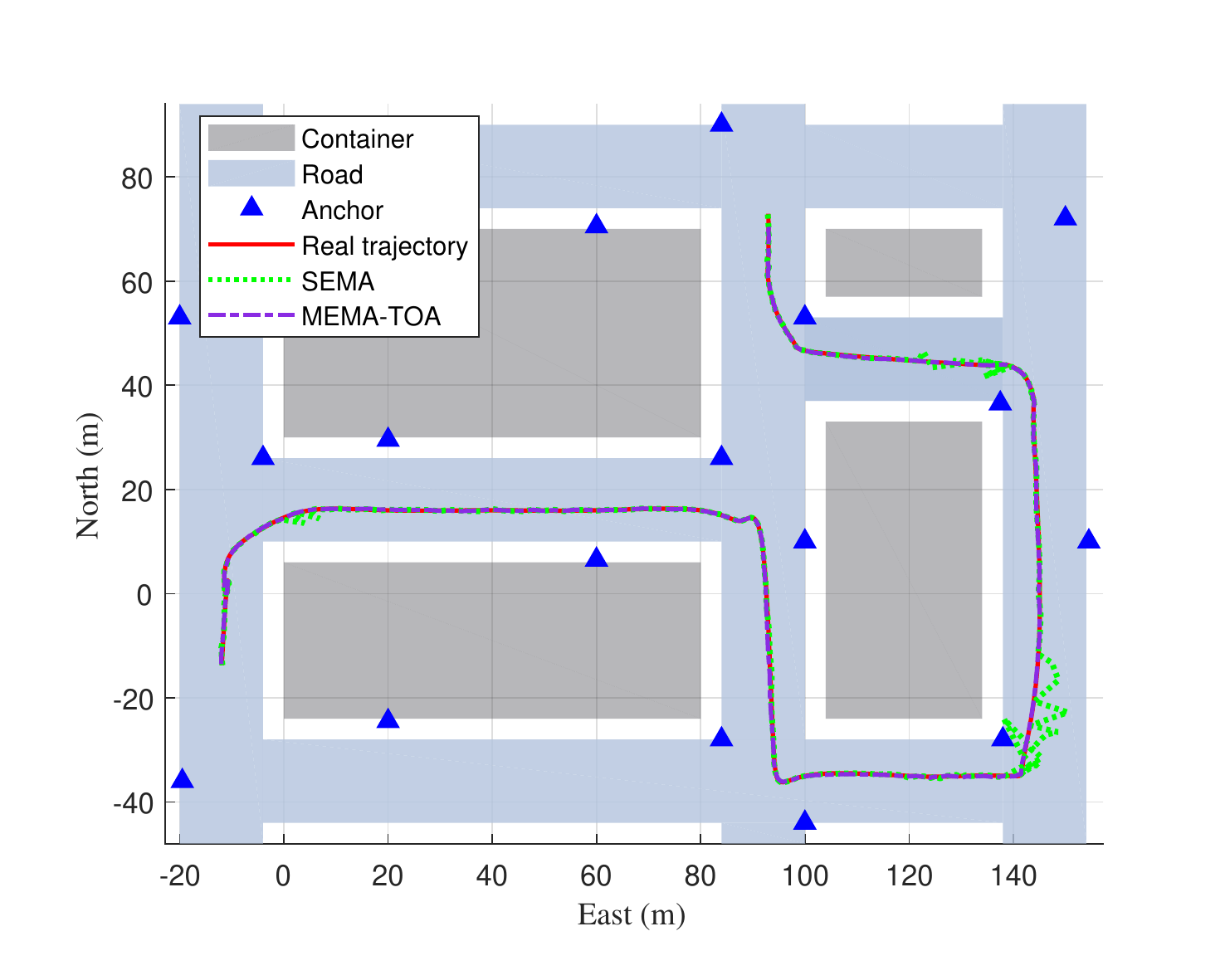}
	\vspace{-1cm}
	\caption{Estimated trajectory in the simulated scene of an unmanned cargo port (top view). The trajectory estimated by the conventional SEMA (green line) has some large deviation from the real trajectory. The trajectory estimated by the new MEMA-TOA (purple line) is close to the real trajectory, showing its superior performance.
}
	\label{fig:trajectory}
	\vspace{-0.3cm}
\end{figure}

For this scene, our MEMA-TOA method with $K=3$ epochs and the conventional SEMA are applied to estimate the position and attitude of the vehicle for each epoch. We initialize the conventional SEMA with a uniformly distributed random guess for the east and north coordinates from the entire area formed by the anchors. The trajectories estimated by the new MEMA-TOA and the conventional SEMA are shown in Fig. \ref{fig:trajectory}. The real trajectory is represented by the red line. The trajectory estimated by the conventional SEMA has some spikes, which indicate that the SEMA converges to erroneous results or ambiguous locations at some epochs. For the areas near the edges and corners where it is difficult to have an evenly distributed anchor geometry, the new MEMA-TOA method achieves good positioning results compared with the conventional SEMA, as shown by the figure. As shown by Fig. \ref{fig:trajectory}, the trajectory estimated by MEMA-TOA is very close to the real one, such that they can hardly be distinguished in the figure. 
This result shows that the new MEMA-TOA method successfully eliminates the ambiguous locations and achieves high-precision positioning results. 

\begin{figure}
	\centering
	\includegraphics[width=0.99\linewidth]{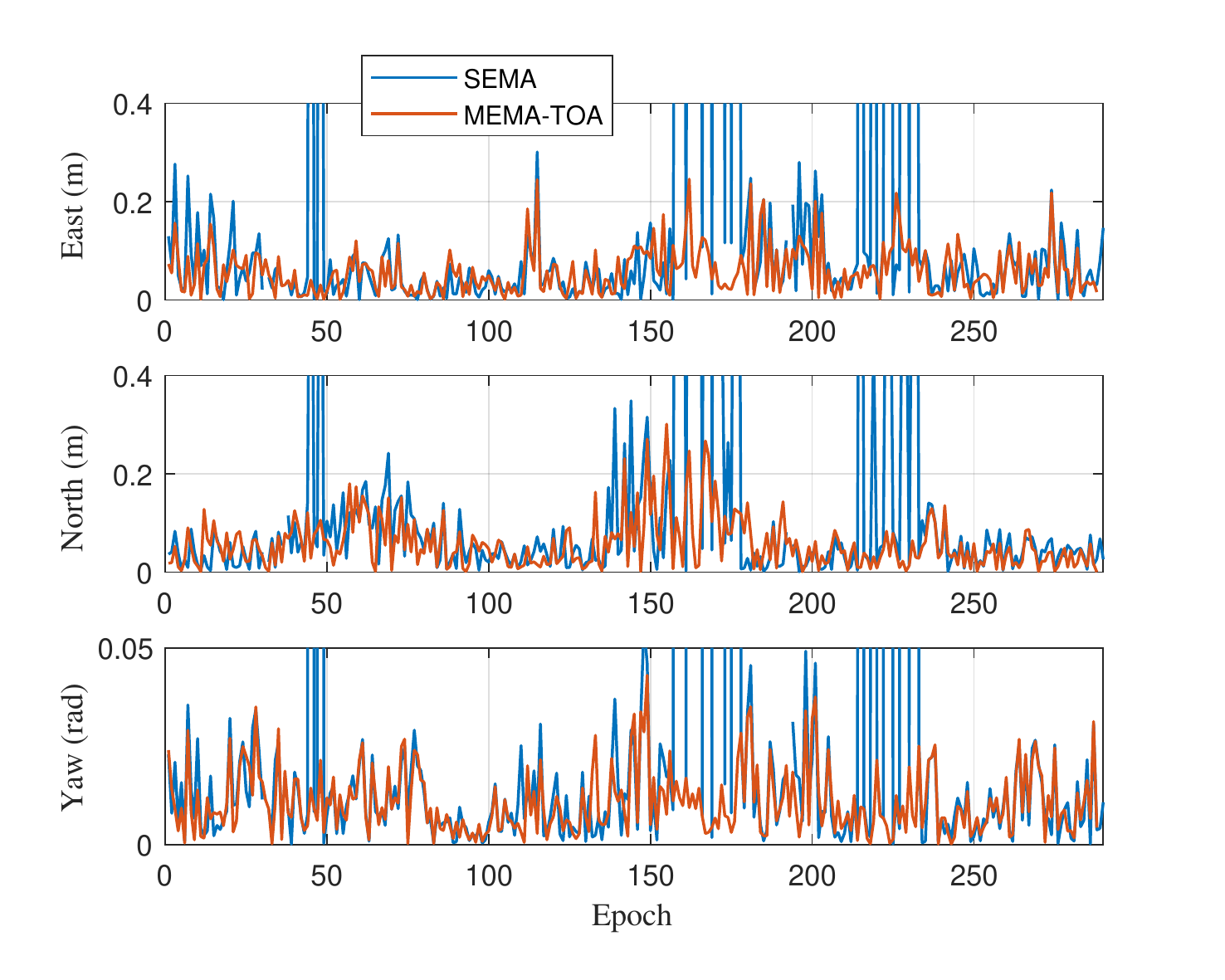}
	\vspace{-0.6 cm}
	\caption{Position and yaw angle estimation error of the new MEMA-TOA method in the unmanned port simulation scene. The estimation error of the new MEMA-TOA method is at decimeter level for the position and lower than 0.05 rad for the yaw angle, both smaller than that of the conventional SEMA method.}
	\label{fig:error_port}
	\vspace{-0.3cm}
\end{figure}

Fig. \ref{fig:error_port} shows the estimation error from the new MEMA-TOA method throughout all epochs. The errors of the convectional SEMA method are also depicted for comparison. For the conventional SEMA method, the errors at 25 epochs are far greater than that of the new MEMA-TOA method as shown in the figure. They are caused by the ambiguous location results. In addition, there are another 6 epochs, at which the conventional iterative SEMA method does not converge since the initial guess is far from the real position. As presented in the figure, compared with the conventional SEMA method,
the new MEMA-TOA method provides accurate and robust positioning results for the whole trajectory, in which the east position, north position and yaw angle errors are lower than 0.3 m, 0.3 m and 0.05 rad, respectively.  

In addition, it should be noted that although three antennas are employed in this simulation, our method is not limited to this number. In practice, the number of antennas can be selected according to the environment and the deployment of anchors.

\section{Conclusion}\label{conclusion}
Autonomous high-precision and robust positioning for vehicles are desperately needed in unmanned warehouses, cargo ports, and similar environments, in which radio broadcast positioning systems are widely employed. However, obstacles in the environments and on the vehicle itself may reduce the number of available signals, and thus degrade the positioning accuracy and robustness. The conventional SEMA method only utilizes measurements from a single epoch, and has the problems of parameter initialization and location ambiguity.

 
In this article, we propose a new MEMA-TOA method. We first develop a new MEMA-SDP to obtain a good initialization. MEMA-SDP provides a global optimum to the approximated convex problem of the original MEMA positioning problem. We then refine the initial positioning result from MEMA-SDP with an iterative algorithm. We take advantage of the MEMA TOAs bridged by inter-epoch constraints, which bring sufficient measurements and improve the geometry of the observed anchors to remove the location ambiguity. In brief, the proposed MEMA-TOA method can obtain high-precision positioning results without a priori information of the initial value. Therefore, it can be used not only as a stand-alone positioning method but also as an initialization, after which other methods such as the EKF can then be adopted in the subsequent epochs. In addition, we derive the CRLB of the MEMA positioning problem and theoretically show that the estimation accuracy of the new MEMA-TOA method is higher than that of the conventional SEMA method. 

Simulation results demonstrate that the MEMA-SDP method successfully provides a good initialization without location ambiguity, and the new MEMA-TOA method has higher position and attitude estimation accuracy and robustness than the conventional SEMA method in harsh environments. All the numerical results are consistent with the theoretical analysis and show the feasibility of the new MEMA-TOA method in real-world intelligent transportation systems and applications.

    
\appendices
\section{Vectorization of Rotation Matrix}\label{Appendix_VEC}
The rotation matrix from frame $ \mathrm{b} $ to frame $ \mathrm{n} $ is \cite{Farrell2008Aided}
\begin{align}
\bm{R} = \left[ {\begin{matrix}
			{\mathrm{c}_{\psi}\mathrm{c}_{ \gamma} }&{\mathrm{s}_ {\psi}\mathrm{c}_{\phi} + \mathrm{c}_{ \psi}\mathrm{s}_{ \gamma} \mathrm{s}_{ \phi} }&{ - \mathrm{s}_{ \psi}\mathrm{s}_{\phi}  + \mathrm{c}_{ \psi}\mathrm{s}_{ \gamma} \mathrm{c}_{\phi}} \\ 
			{\mathrm{s}_ {\psi}\mathrm{c}_{ \gamma} }&{ - \mathrm{c}_ {\psi}\mathrm{c}_ {\phi}  + \mathrm{s} _{\psi}\mathrm{s}_{ \gamma} \mathrm{s}_ {\phi} }&{\mathrm{c}_{ \psi}\mathrm{s}_ {\phi } + \mathrm{s}_ {\psi}\mathrm{s} _{\gamma} \mathrm{c}_ {\phi} } \\ 
			{\mathrm{s}_ {\gamma} }&{ - \mathrm{c}_ {\gamma }\mathrm{s}_ {\phi} }&{ - \mathrm{c}_{ \gamma }\mathrm{c}_{ \phi} } 
	\end{matrix}} \right]\text{,}
\end{align}
where $ \gamma $, $ \phi $ and $\psi $  are the known roll angle, the known pitch angle and the unknown yaw angle, respectively. $\mathrm{s}_{\phi}$, $\mathrm{s}_{\gamma}$, $\mathrm{s}_{\psi}$, $\mathrm{c}_{\phi}$, $ \mathrm{c}_{\gamma}$ and $\mathrm{c}_{\psi}$ represent $ \operatorname{sin}\phi$, $\operatorname{sin}\gamma$, $\operatorname{sin}\psi$, $ \operatorname{cos}\phi$, $\operatorname{cos}\gamma$ and $\operatorname{cos}\psi$, respectively.

For epoch $ k $, $ \bm{R}_{(k)} $ is a function of $ {\psi _{\left( k \right)}} $, and can be vectorized as 
\begin{equation}
\operatorname{vec}\left( {{{\bm{R}}_{\left( k \right)}}} \right) = \left[ {\begin{matrix}
			{\mathrm{c}_ {\psi _{\left( k \right)}}\mathrm{c}_ {\gamma }} \\ 
			{\mathrm{s}_ {\psi _{\left( k \right)}}\mathrm{c}_ {\gamma} } \\ 
			{\mathrm{s}_{ \gamma }} \\ 
			{\mathrm{s}_ {\psi _{\left( k \right)}}\mathrm{c}_{ \phi } + \mathrm{c}_ {\psi _{\left( k \right)}}\mathrm{s}_{ \gamma} \mathrm{s}_{ \phi} } \\ 
			{ - \mathrm{c}_ {\psi _{\left( k \right)}}\mathrm{c}_{ \phi}  + \mathrm{s}_ {\psi _{\left( k \right)}}\mathrm{s}_{ \gamma} \mathrm{s}_{ \phi} } \\ 
			{ - \mathrm{c} _{\gamma }\mathrm{s}_{ \phi} } \\ 
			{ - \mathrm{s}_ {\psi _{\left( k \right)}}\mathrm{s}_{ \phi } + \mathrm{c}_ {\psi _{\left( k \right)}}\mathrm{s} _{\gamma} \mathrm{c}_{ \phi} } \\ 
			{\mathrm{c}_ {\psi _{\left( k \right)}}\mathrm{s}_{ \phi } + \mathrm{s}_ {\psi _{\left( k \right)}}\mathrm{s}_ {\gamma} \mathrm{c}_{ \phi} } \\ 
			{ - \mathrm{c}_ {\gamma} \mathrm{c}_ {\phi} } 
				\end{matrix}} \right]\text{.}
\end{equation}

Utilizing  $ {{\bm{u}}_{\left( k \right)}}$, 
$\bm{\Gamma }$ and $ \bm{\alpha }$ defined in Section \ref{TOASDP},
we then have
\begin{equation}
\operatorname{vec}({\bm{R}_{\left( k \right)}}) = {\bm{\alpha }} + {\bm{\Gamma }}{\bm{u}_{\left( k \right)}}\text{.}
\end{equation}

\section{ Linearization in Step 2}\label{Appendix_ME}
We conduct Taylor series expansion of the TDOA equation in (\ref{eq:deltarhoij}) at $\hat {\bm{\Theta}} $, ignore the higher-order terms, and come to a collective form of the linearized $ K $-epoch TDOA equation as  
\begin{equation}\label{eq:linear_TOA}
	\delta {\bm{z}_{\mathrm{TOA}}} = {\bm{H}_{\mathrm{TOA}}}\cdot\delta \bm{\Theta}+ {{\bm{\varepsilon}}_{\mathrm{TOA}}}\text{,}
\end{equation}
where  the increment $\delta \bm{\Theta} = \bm{\Theta} - \hat {\bm{\Theta} }$, 
\begin{equation}
\nonumber	
\begin{aligned}
     \delta {\bm{z}_{\mathrm{TOA}}}& = \left[ {
			{\delta \bm{z}_{\left( { 1} \right)}^T},\ \cdots ,\ {\delta \bm{z}_{\left( { K} \right)}^T} 
} \right]^T\text{,} \\
	 {{\bm{H}}_{\mathrm{TOA}}} &= \operatorname{blkdiag}\left(\bm{H}_{\left( 1 \right)}, \dots,{{{\bm{H}}_{(K)}}} 
	 \right)\text{,}\\
	 \bm{Q} _{ \bm{\varepsilon}_{\mathrm{TOA}}}&=\operatorname{blkdiag}		\left(\bm{Q}_{\Delta\varepsilon_{(1)}},\dots,\bm{Q}_{\Delta\varepsilon_{(K)}}\right),
\end{aligned}
\end{equation}
where $\delta \bm{z}_{(k)} = \Delta\bm{\rho}_{(k)} - \Vert\bm{p}_{(k)}^{(j)}-\bm{p}_{i(k)}\Vert-\Vert\bm{p}_{(k)}^{(1)}-\bm{p}_{1(k)}\Vert $, $\bm{Q}_{\Delta\varepsilon_{(k)}}$ is derived in (\ref{eq:deltarho_vecter}), and  $ \bm{H}_{( k )} = \left. \frac{\partial \bm{g}_{( k )}{(\bm{\theta}_{(k)})}}{\partial \bm{\theta}_{( k )}} \right|_{\hat {\bm{\theta}}_{( k )}}$, in which $\bm{g}_{( k )}$ is defined in \eqref{eq:deltarho_vecter}.

For the inter-epoch change constraints, we expand the equation in (\ref{eq:odon}) at  $\hat {\bm{\Theta}} $, eliminate the higher-order term, and have the $K$-epoch  inter-epoch constraint equation
as
\begin{equation} \label{eq:linear_IP}
	\delta {\bm{z}_{{\mathrm{IP}}}} = {\bm{H}_{{\mathrm{IP}}}}\cdot\delta \bm{\Theta} + {{\bm{\varepsilon}}_{{\mathrm{IP}}}}\text{,}
\end{equation}
where
\begin{align}
 \nonumber{{\bm{H}}_{{\mathrm{IP}}}}& = {\left[ {
			{{\bm{H}}_{{\mathrm{IP}}\left( 2 \right)}^T},\ \cdots ,\ {{\bm{H}}_{{\mathrm{IP}}\left( K \right)}^T} 
	} \right]^T}\text{,}\\
\nonumber	\delta \bm{z}_{\mathrm{IP}}&=\left[ 
		\delta \bm{z}_{\mathrm{IP}\left( 2 \right)}^T,\ \cdots,\ {\delta \bm{z}_{\mathrm{IP}\left( K \right)}^T} 
 \right]^T\text{,}\\
	\nonumber\bm{Q}_{\bm{\varepsilon}_{\mathrm{IP}}}&= \operatorname{blkdiag}\left(
	\bm{Q}_{\bm{\varepsilon}_{\mathrm{IP}(2)} },\cdots,\bm{Q}_{ \bm{\varepsilon}_{\mathrm{IP}(K)} }\right),
	\end{align}	
where $\delta \bm{z}_{{\mathrm{IP}}\mathrm( k )}= \Delta\tilde{\bm{\theta}}_{(k,k-1)}^{\mathrm{b}_{(k-1)}}- \bm{g}_{\mathrm{IP}( k )}\left(\bm{\theta}_{(k)},\bm{\theta}_{(k-1)}\right)$, the covariance matrix $\bm{Q}_{ \bm{\varepsilon}_{\mathrm{IP}(k)}}$ is $  \operatorname{diag}\left( {\sigma _p^2},{\sigma _p^2},{\sigma _\psi ^2} \right) $, and 
\begin{align}
\nonumber &{\bm{H}_{\mathrm{IP}\left( k \right)}} \\
 \nonumber &=\underbrace {\left[ {\begin{matrix}
			{{{\bm{0}}_{3 \times 3}}}& \ldots & {\left. {\frac{{\partial {\bm{g}_{\mathrm{IP}\left( k \right)}}}}{{\partial \bm{\theta}_{(k-1)}}}} \right|_{\hat{\bm {\theta}}_{(k-1)}}}&{\left. {\frac{{\partial {\bm{g}_{\mathrm{IP}\left( k \right)}}}}{{\partial \bm{\theta}_{(k)}}}} \right|_{\hat{\bm {\theta}}_{(k)}}}& \cdots &{{{\bm{0}}_{3 \times 3}}} 
	\end{matrix}} \right]}_{3 \times 3K}\text{,}
\end{align} 
in which, $\bm{g}_{\mathrm{IP}\left( k \right)}$ is defined in (\ref{eq:rewrite_odo}).

\section{CRLB for MEMA Problem}\label{Appendix_CRLB}
We derive the CRLB of the MEMA-TOA positioning problem. 

According to the noise model given by (\ref{eq:rhoij}) and (\ref{eq:odo1}),  all the noises of MEMA TOA measurements and the inter-epoch change constraints are independent Gaussian noises. Utilizing the linearized equations \eqref{eq:linear_TOA} and \eqref{eq:linear_IP} in Appendix \ref{Appendix_ME}, the  Fisher information matrix for MEMA problem is 
	\begin{align}\label{eq:F_MEMA}
\nonumber	\mathtt {F}_{\mathrm{MEMA}}=&   
	{\bm{H}}_{\mathrm{TOA}}^T\bm{W}_{\mathrm{TDOA}}{{\bm{H}}_{\mathrm{TOA}}}	+ {\bm{H}}_{\mathrm{IP}}^T\bm{W}_{\mathrm{IP}}{{\bm{H}}_{\mathrm{IP}}}\\
	=&\mathtt{F}_{\mathrm{MEMA,TOA}} + {\bm{H}}_{\mathrm{IP}}^T\bm{W}_{\mathrm{IP}}{{\bm{H}}_{\mathrm{IP}}}\text{,}	
\end{align}
where $ \bm{W}_{\mathrm{TOA}}=\bm{Q} _{ \bm{\varepsilon}_{\mathrm{TOA}} }^{-1} $, $ \bm{W}_{\mathrm{IP}}=\bm{Q}_{ \bm{\varepsilon}_{\mathrm{IP}} }^{-1} $, and $\mathtt{F}_{\mathrm{MEMA,TOA}}$ is the Fisher information matrix derived without inter-epoch constraints.	

	Utilizing the knowledge of matrix analysis \cite{Horn2012matrix}, we have
	\begin{align}\label{eq:FIM_ME_2}
		\mathtt{F}_\mathrm{MEMA}^{- 1} =&\left( \mathtt{F}_{\mathrm{MEMA,TOA}} + {\bm{H}}_{\mathrm{IP}}^T\bm{W}_{\mathrm{IP}}{{\bm{H}}_{\mathrm{IP}}} \right)^{ - 1}  \\
		\nonumber	=& \mathtt{F}_{\mathrm{MEMA,TOA}}^{ - 1} \\
	\nonumber	&- \mathtt{F}_{\mathrm{MEMA,TOA}}^{ - 1}{\bm{H}}_{\mathrm{IP}}^T\bm{\Xi}{\bm{H}}_{\mathrm{IP}}\mathtt{F}_{\mathrm{MEMA,TOA}}^{-1} \text{,}
	\end{align}
	where  $ \bm{\Xi}\triangleq  \bm{W}_{\mathrm{IP}}^{-1 }+ \bm{H}_{\mathrm{IP}}\mathtt{F}_{\mathrm{MEMA,TOA}}^{-1}{\bm{H}}_{\mathrm{IP}}^{T} $.
	
	According to the definition of $ \mathtt{F}_{\mathrm{MEMA,TOA}}$,  it is a block diagonal matrix. Each block of $ \mathtt{F}_{\mathrm{MEMA,TOA}}$ is a symmetric positive definite matrix, and so is $\mathtt{F}_{\mathrm{MEMA,TOA}}^{ - 1} $. We conduct eigenvalue decomposition on $ \mathtt{F}_{\mathrm{MEMA,TOA}}^{ - 1} $ as
	\begin{equation}\label{eq:FIM_ME_ED}
		\mathtt{F}_{\mathrm{MEMA,TDOA}}^{- 1} = \bm{Z}^T\bm{\Lambda }\bm{Z}\text{,}
	\end{equation}
	in which $ \bm{\Lambda }=\operatorname{diag}\left( 	\lambda_{1},\cdots,\lambda _{3K} 	 \right) $ is the diagonal matrix constructed by the positive eigenvalues of   $ \mathtt{F}_{\mathrm{MEMA,TOA}}^{ - 1} $, and $ \bm{Z}$   is the corresponding orthogonal matrix.
	
	Let $  {\bm{H}}_{\mathrm{IP}}\bm{Z}^T = \left[ 
		\bm{a}_1,\ \dots,\ \bm{a}_{3K}  \right]  $, in which $ \bm{a}_l\; (l=1,\dots,3K )$ are column vectors. Then,
	\begin{align}\label{eq:HFIMG_ED}
		\nonumber{\bm{H}}_{\mathrm{IP}}\mathtt{F}_{\mathrm{MEMA,TOA}}^{- 1}{\bm{H}}_{\mathrm{IP}}^{T} &= {\bm{H}}_{\mathrm{IP}}\bm{Z}^T{\bm{\Lambda }}\bm{Z}{\bm{H}}_{\mathrm{IP}}^{T} \\
		&= \sum\limits_{l = 1}^{3K} \lambda _l\bm{a}_l\bm{a}_l^T
	\end{align}
	indicates that $ {\bm{H}}_{\mathrm{IP}}\mathtt{F}_{\mathrm{MEMA,TOA}}^{- 1}{\bm{H}}_{\mathrm{IP}}^{T}  $ is a symmetric positive semidefinite matrix.
	
	 Conduct the eigenvalue decomposition
	\begin{equation}
		{\left( {\bm{H}}_{\mathrm{IP}}\mathtt{F}_{\mathrm{MEMA,TOA}}^{- 1}{\bm{H}}_{\mathrm{IP}}^{T}  \right)^{ - 1}} = {\bm{M}}_1^T{{\bm{\Lambda }}_1}{{\bm{M}}_1}\text{,}
	\end{equation}
	where $\nonumber \bm{\Lambda }_{1}=\operatorname{diag}\left( 	\lambda_{1,1},\cdots ,\lambda_{1,L_K} \right)$
	   is constructed by the non-negative eigenvalues,  $ \bm{M}_1 $ is the corresponding orthogonal matrix, $ L_K $ is the number of TDOA measurements for $ K $ epochs, where $ L_K = \sum_{k=1}^K L_{(k)}$ and $ L_{(k)} =\sum_{n = 1}^N M_{i\left( k \right)}-1$.
	   
	    Consequently,
	\begin{equation}\label{eq:B}
		\bm{\Xi} = {\bm{M}}_1^T{\left( {{{\bm{\Lambda }}_2}+ {{\bm{\Lambda }}_1}} \right)^{ - 1}}{{\bm{M}}_1}\text{,}
	\end{equation}
	where $\bm{\Lambda }_{2}=\operatorname{diag}\left(	\lambda_{2,1},\cdots,\lambda_{2,L_K} 	 \right)$ is constructed by the negative eigenvalues of $ \bm{W}_{\mathrm{IP}} $.

	Let 
	\begin{equation}
		 \bm{A}=\mathtt{F}_{\mathrm{MEMA,TOA}}^{ - 1}{\bm{H}}_{\mathrm{IP}}^{T}\bm{\Xi}{\bm{H}}_{\mathrm{IP}}\mathrm{F}_{\mathrm{MEMA,TOA}}^{-1} \text{,}
		 \end{equation}
	 we have 
	 \begin{align}\label{eq:FIM_ME_3}
	 	\mathtt{F}_\mathrm{MEMA}^{- 1} =\mathtt{F}_{\mathrm{MEMA,TOA}}^{ - 1} - \bm{A}\text{.}
	 \end{align}
 
	 Substituting (\ref{eq:B}) into $ \bm{A} $ gives
	\begin{align}
		&{\bm{A}} = \\
	\nonumber	&{\mathtt{F}}_{\mathrm{MEMA,TOA}}^{- 1}{\bm{H}}_{\mathrm{IP}}^{T}{\bm{M}}_1^T{\left( {{\bm{\Lambda }}_2} + {{\bm{\Lambda }}_1} \right)^{ - 1}}{{\bm{M}}_1}{{\bm{H}}_{\mathrm{IP}}}{\mathtt{F}}_{\mathrm{MEMA,TOA}}^{ - 1}\text{.}
	\end{align}   
 
	Let 
	\begin{equation}
	\mathtt{F}_{\mathrm{MEMA,TOA}}^{ - 1}{\bm{H}}_{\mathrm{IP}}^{T}{\bm{M}}_1^T = \left[ 
			{{{\bm{b}}_1}},\ \cdots ,\ {{{\bm{b}}_{L_K}}} 
\right] \text{.}
\end{equation}

 Similar to (\ref{eq:HFIMG_ED}), $ \bm{A} $ can be rewritten as
	\begin{equation}
		{\bm{A}} = \sum\limits_{l = 1}^{L_K} {{{\left( { {\lambda _{2,l}} + {\lambda _{1,l}}} \right)}^{ - 1}}} {{\bm{b}}_l}{\bm{b}}_l^T \text{.}
	\end{equation}
	
	According to the definition of $ \bm{W}_{\mathrm{IP}} $ and $ \bm{\Lambda }_{2}$, $ {\lambda _{2,l}} > 0 $. Moreover, since $ {\lambda _{1,l}} \geqslant 0 $, $ \bm{A} $ is also a symmetric positive semidefinite matrix, in which the diagonal elements are non-negative.
	
	We denote the $v$-th diagonal element of $\mathtt{F}_{\mathrm{MEMA}} $, $ \mathtt{F}_{\mathrm{MEMA,TOA}} $ and $ \bm{A}  $ by $ d_{{\mathrm{MEMA}},v}$, $d_{{\mathrm{MEMA,TOA}},v}$ and $ d_{A,v}$. According to the properties of the positive semidefinite matrix, the diagonal elements are all non-negative. Therefore, based on (\ref{eq:FIM_ME_3}), we obtain the inequality as
	\begin{equation}\label{eq:d}
		d_{\mathrm{MEMA},v} = d_{{\mathrm{MEMA,TOA}},v} - {d_{A,v}} \leqslant d_{{\mathrm{MEMA,TOA}},v} \text{.}
	\end{equation}
	
	According to (\ref{eq:CRLB}), we have
	\begin{equation}
		{\rm{CRLB}}_{\mathrm{MEMA},v} \leqslant {\rm{CRLB}}_{\mathrm{MEMA,TOA},v} \text{.}
	\end{equation}

Note that according to the definition of $\mathtt{F}_{\mathrm{MEMA,TOA}}$ in \eqref{eq:F_MEMA} and ${\mathtt{F}}_{\mathrm{SEMA}(k)}$ in \eqref{eq:FIM_SE}, we have
\begin{equation}\label{eq:FIM_TDOA}
	\mathtt{F}_{\mathrm{MEMA,TOA}} = 
\operatorname{blkdiag}\left( 
			{\mathtt{F}}_{\mathrm{SEMA}(1)},\cdots,{\mathtt{F}}_{\mathrm{SEMA}( K )} 
\right) \text{.}
\end{equation}
 $ \mathtt{F}_{\mathrm{MEMA,TOA}}$ is a block diagonal matrix and its inverse is equal to the block diagonal matrix formed by the inverse of each blocks. Consequently,
 \begin{equation}
 	{\rm{CRLB}}_{{\mathrm{MEMA}},v} \leqslant {\rm{CRLB}}_{{\mathrm{SEMA}},v} \text{.}
 \end{equation}
 
The theoretical estimation error of the new MEMA-TOA is lower than that of the conventional SEMA, i.e. a higher positioning accuracy can be achieved by the new MEMA-TOA method.


%


\ifCLASSOPTIONcaptionsoff
    \newpage
\fi



\bibliographystyle{IEEEtran}
\bibliography{IEEEabrv,paper_an}
\end{document}